\documentclass{vgtc}                          

\graphicspath{{figures/}{pictures/}{images/}{./}} 

\usepackage{times}                     

\usepackage{tabu}                      
\usepackage{booktabs}                  

\usepackage{mathptmx}                  

\usepackage{glossaries}
\usepackage{tabularx} 
\usepackage{array}
\newcolumntype{Y}{>{\centering\arraybackslash}X}
\usepackage[nolist]{acronym}
\usepackage[inline]{enumitem}

\vgtcinsertpkg

\usepackage{tikz}
\newcommand{\xmark}{%
\tikz[scale=0.23] {
    \draw[line width=0.7,line cap=round] (0,0) to [bend left=6] (1,1);
    \draw[line width=0.7,line cap=round] (0.2,0.95) to [bend right=3] (0.8,0.05);
}}
\newcommand{\cmark}{%
\tikz[scale=0.23] {
    \draw[line width=0.7,line cap=round] (0.25,0) to [bend left=10] (1,1);
    \draw[line width=0.8,line cap=round] (0,0.35) to [bend right=1] (0.23,0);
}}

\title{Multi-Layer Gaussian Splatting for Immersive Anatomy Visualization}

\author{
Constantin Kleinbeck\thanks{e-mail: constantin.kleinbeck@tum.de}\\\parbox{3.1in}{\scriptsize Technical University of Munich, Clinic for Orthopedics and Sports Orthopedics, TUM University Hospital, Munich, Germany}%
\and Hannah Schieber\thanks{e-mail: hannah.schieber@tum.de}\\\parbox{3.1in}{\scriptsize Technical University of Munich, Clinic for Orthopedics and Sports Orthopedics, TUM University Hospital, Munich, Germany}\\%
\and Klaus Engel\thanks{e-mail: engel.klaus@siemens-healthineers.com}\\\scriptsize Siemens Healthineers AG, Erlangen, Germany%
\and Ralf Gutjahr\thanks{e-mail: ralf.gutjahr@siemens-healthineers.com}\\\scriptsize Siemens Healthineers AG, Forchheim, Germany\\%
\and Daniel Roth\thanks{e-mail: daniel.roth@tum.de}\\\scriptsize Technical University of Munich, Clinic for Orthopedics and Sports Orthopedics, TUM University Hospital, Munich, Germany%
}%

\teaser{
  \centering
  \includegraphics[width=\linewidth, alt={Teaser figure showing how a layered Gaussian Splatting representation is created from multiple input images, rendered with multiple transfer functions showing different parts of the anatomy and interacted with in VR.}]{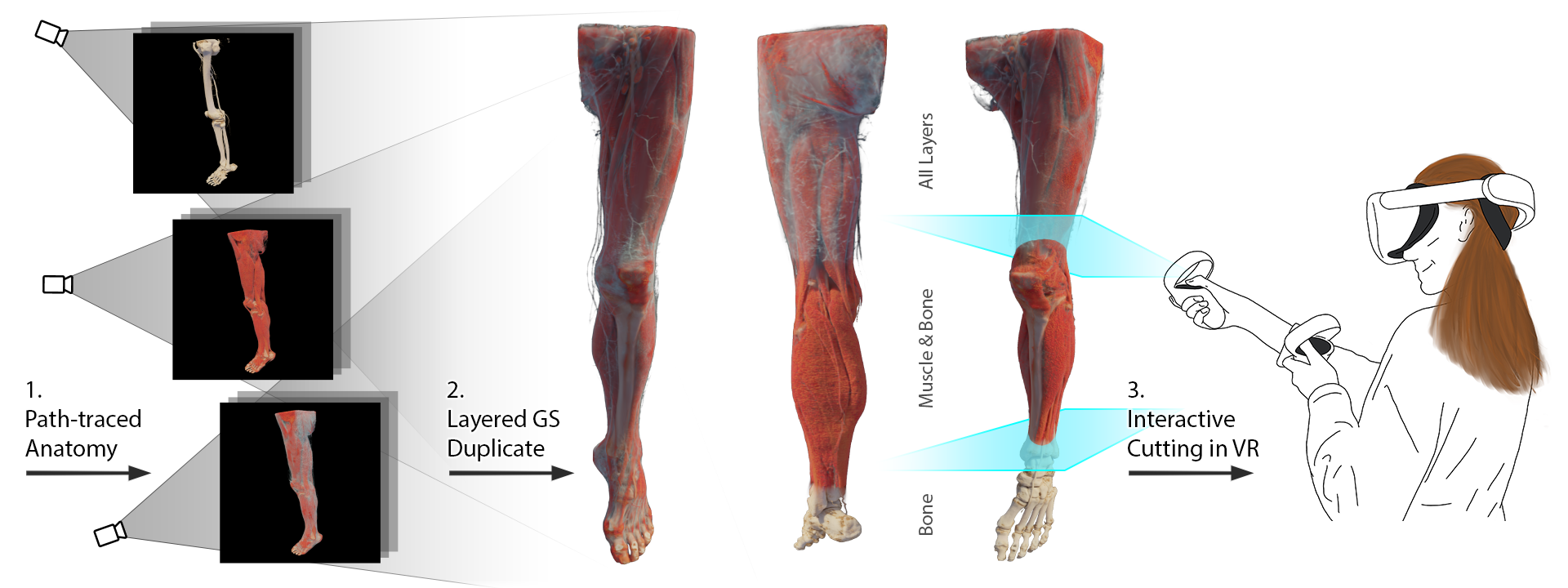}
  \caption{%
  	Our layered gaussian splatting (GS) model is consecutively trained on multiple renderings of the same computed tomography (CT) scan, visualizing different parts of the anatomy. It can be viewed in real-time in virtual reality (VR), and the trained anatomies can selectively be shown and cut to create engaging visuals and bridge the interactivity gap to full path tracing.%
  }
  \label{fig:teaser}
}

\abstract{%
In medical image visualization, path tracing of volumetric medical data like \ac{ct} scans produces lifelike three-dimensional visualizations. Immersive \ac{vr} displays can further enhance the understanding of complex anatomies. Going beyond the diagnostic quality of traditional 2D slices, they enable interactive 3D evaluation of anatomies, supporting medical education and planning. Rendering high-quality visualizations in real-time, however, is computationally intensive and impractical for compute-constrained devices like mobile headsets.

We propose a novel approach utilizing \ac{gs} to create an efficient but static intermediate representation of CT scans. We introduce a layered GS representation, incrementally including different anatomical structures while minimizing overlap and extending the GS training to remove inactive Gaussians. We further compress the created model with clustering across layers. 

Our approach achieves interactive frame rates while preserving anatomical structures, with quality adjustable to the target hardware. Compared to standard GS, our representation retains some of the explorative qualities initially enabled by immersive path tracing. Selective activation and clipping of layers are possible at rendering time, adding a degree of interactivity to otherwise static GS models. This could enable scenarios where high computational demands would otherwise prohibit using path-traced medical volumes.}


\begin{document}
\begin{acronym}[Bspwwww.]  

\acro{ar}[AR]{augmented reality}
\acro{ap}[AP]{average precision}
\acro{api}[API]{application programming interface}
\acroplural{ann}[ANN]{artifical neural networks}
\acro{bev}[BEV]{bird eye view}
\acro{rbob}[BRB]{Bottleneck residual block}
\acroplural{rbob}[BRBs]{Bottleneck residual blocks}
\acro{mbiou}[mBIoU]{mean Boundary Intersection over Union}
\acro{cai}[CAI]{computer-assisted intervention}
\acro{ce}[CE]{cross entropy}
\acro{cad}[CAD]{computer-aided design}
\acro{cnn}[CNN]{convolutional neural network}
\acro{ct}[CT]{computed tomography}
\acro{crf}[CRF]{conditional random fields}
\acro{dpc}[DPC]{dense prediction cells}
\acro{dla}[DLA]{deep layer aggregation}
\acro{dnn}[DNN]{deep neural network}
\acroplural{dnn}[DNNs]{deep neural networks}
\acro{dvr}[DVR]{direct volume rendering}
\acro{da}[DA]{domain adaption}
\acro{dr}[DR]{domain randomization}
\acro{fat}[FAT]{falling things}
\acro{fcn}[FCN]{fully convolutional network}
\acroplural{fcn}[FCNs]{fully convolutional networks}
\acro{fov}[FoV]{field of view}
\acro{fv}[FV]{front view}
\acro{fp}[FP]{False Positive}
\acro{fpn}[FPN]{feature Pyramid network}
\acro{fn}[FN]{False Negative}
\acro{fmss}[FMSS]{fast motion sickness scale}
\acro{gan}[GAN]{generative adversarial network}
\acroplural{gan}[GANs]{generative adversarial networks}
\acro{gcn}[GCN]{graph convolutional network}
\acroplural{gcn}[GCNs]{graph convolutional networks}
\acro{gs}[GS]{gaussian splatting}
\acro{hmi}[HMI]{Human-Machine-Interaction}
\acro{hmd}[HMD]{Head Mounted Display}
\acroplural{hmd}[HMDs]{head mounted displays}
\acro{iou}[IoU]{intersection over union}
\acro{irb}[IRB]{inverted residual bock}
\acroplural{irb}[IRBs]{inverted residual blocks}
\acro{ipq}[IPQ]{igroup presence questionnaire}
\acro{knn}[KNN]{k-nearest-neighbor}
\acro{lidar}[LiDAR]{light detection and ranging}
\acro{lsfe}[LSFE]{large scale feature extractor}
\acro{llm}[LLM]{large language model}
\acro{map}[mAP]{mean average precision}
\acro{mc}[MC]{mismatch correction module}
\acro{miou}[mIoU]{mean intersection over union}
\acro{mis}[MIS]{Minimally Invasive Surgery}
\acro{msdl}[MSDL]{Multi-Scale Dice Loss}
\acro{ml}[ML]{Machine Learning}
\acro{mlp}[MLP]{multilayer perception}
\acro{miou}[mIoU]{mean Intersection over Union}
\acro{mri}[MRI]{magnetic resonance imaging}
\acro{nn}[NN]{neural network}
\acroplural{nn}[NNs]{neural networks}
\acro{ndd}[NDDS]{NVIDIA Deep Learning Data Synthesizer}
\acro{nocs}[NOCS]{Normalized Object Coordiante Space}
\acro{nerf}[NeRF]{Neural Radiance Fields}
\acro{NVISII}[NVISII]{NVIDIA Scene Imaging Interface}
\acro{ngp}[NGP]{neural graphics primitives}
\acro{or}[OR]{Operating Room}
\acro{pbr}[PBR]{physically based rendering}
\acro{psnr}[PSNR]{peak signal-to-noise ratio}
\acro{pnp}[PnP]{Perspective-n-Point}
\acro{rv}[RV]{range view}
\acro{roi}[ROI]{region of interest}
\acroplural{roi}[ROIs]{region of interests}
\acro{rbab}[BB]{residual basic block}
\acro{ras}[RAS]{robot-assisted surgery}
\acroplural{rbab}[BBs]{residual basic blocks}
\acro{spp}[SPP]{spatial pyramid pooling}
\acro{sh}[SH]{spherical harmonics}
\acro{sgd}[SGD]{stochastic gradient descent}
\acro{sdf}[SDF]{signed distance function}
\acro{sfm}[SfM]{structure-from-motion}
\acro{sam}[SAM]{Segment-Anything}
\acro{sus}[SUS]{system usability scale}
\acro{ssim}[SSIM]{structural similarity index measure}
\acro{sfm}[SfM]{structure from motion}
\acro{slam}[SLAM]{simultaneous localization and mapping}
\acro{tp}[TP]{True Positive}
\acro{tn}[TN]{True Negative}
\acro{thor}[thor]{The House Of inteRactions}
\acro{tsdf}[TSDF]{signed distance function}
\acro{vr}[VR]{Virtual Reality}
\acro{ycb}[YCB]{Yale-CMU-Berkeley}

\acro{ar}[AR]{augmented reality}
\acro{ate}[ATE]{absolute trajectory error}
\acro{bvip}[BVIP]{blind or visually impaired people}
\acro{cnn}[CNN]{convolutional neural network}
\acro{c2f}[c2f]{coarse-to-fine}
\acro{fov}[FoV]{field of view}
\acro{gan}[GAN]{generative adversarial network}
\acro{gcn}[GCN]{graph convolutional Network}
\acro{gnn}[GNN]{Graph Neural Network}
\acro{hmi}[HMI]{Human-Machine-Interaction}
\acro{hmd}[HMD]{head-mounted display}
\acro{mr}[MR]{mixed reality}
\acro{iot}[IoT]{internet of things}
\acro{llff}[LLFF]{Local Light Field Fusion}
\acro{bleff}[BLEFF]{Blender Forward Facing}

\acro{lpips}[LPIPS]{learned perceptual image patch similarity}
\acro{nerf}[NeRF]{neural radiance fields}
\acro{nvs}[NVS]{novel view synthesis}
\acro{mlp}[MLP]{multilayer perceptron}
\acro{mrs}[MRS]{Mixed Region Sampling}

\acro{or}[OR]{Operating Room}
\acro{pbr}[PBR]{physically based rendering}
\acro{psnr}[PSNR]{peak signal-to-noise ratio}
\acro{pnp}[PnP]{Perspective-n-Point}
%
\acro{sus}[SUS]{system usability scale}
\acro{ssim}[SSIM]{similarity index measure}
\acro{sfm}[SfM]{structure from motion}
\acro{slam}[SLAM]{simultaneous localization and mapping}

\acro{tp}[TP]{True Positive}
\acro{tn}[TN]{True Negative}
\acro{thor}[thor]{The House Of inteRactions}
\acro{ueq}[UEQ]{User Experience Questionnaire}
\acro{vr}[VR]{virtual reality}
\acro{who}[WHO]{World Health Organization}
\acro{xr}[XR]{extended reality}
\acro{ycb}[YCB]{Yale-CMU-Berkeley}
\acro{yolo}[YOLO]{you only look once}
\end{acronym}


\firstsection{Introduction}

\maketitle

\Ac{ct} scanning is a widely used imaging technique that provides exceptionally high-resolution volumetric data, allowing detailed examinations of the human body and its complex structures. Traditionally, clinicians and radiologists rely on 2D slice views of such volumetric data for diagnosis and treatment planning. Yet, 3D visualization techniques have been a research subject for around four decades~\cite{frieder_back--front_1985, levoy_display_1988}. Today, they are increasingly finding their way into medical applications in fields like education, training, surgical planning, and decision-making~\cite{zhou_review_2022}. 

Volumetric path tracing is a modern alternative to traditional ray marching for 3D medical data visualization~\cite{engel_real-time_2006, comaniciu_shaping_2016}. This computationally intensive technique models photons' path and interactions, producing realistic renderings, including improved lighting and shadows. The resulting visual fidelity particularly benefits medical education and training~\cite{binder_leveraging_2019, binder_cinematic_2021}, making complex anatomical structures more comprehensible. When presented on immersive headsets, the stereoscopic environment clearly conveys spatial relationships and context. This can create an engaging, explorative learning environment. 

Direct volume rendering using ray marching has been a longstanding technique for visualizing volumetric medical data~\cite{engel_real-time_2006}. While effective, it often struggles to produce realistic renderings. Volumetric path tracing, a technique borrowed from the special effects industry, has emerged as a promising alternative \cite{comaniciu_shaping_2016}. This method invests more computing time to model photons' interactions and path through the scene. Rendered medical volumes boast better physical properties, including realistic lighting and shadows, enhancing the overall visual fidelity. This enhanced realism is particularly beneficial for educational and training purposes \cite{binder_leveraging_2019}, by making complex anatomical structures easier to understand~\cite{binder_cinematic_2021}. Immersive 3D displays can present this inherently three-dimensional data in a stereoscopic environment. This makes relationships and context readily apparent, providing an engaging and explorative learning environment.

Yet, these advancements come with their own set of challenges. With their increasing resolution, modern \ac{ct} scanners can produce file sizes in the hundreds to thousands of MB, with specialized devices getting into tens of GB per scan~\cite{walsh_imaging_2021}. While not problematic for workstations, distribution to and storage on mobile devices can quickly get complicated. Path tracing, while capable of producing high-quality outputs, requires substantial computational power, especially for real-time rendering. This presents a particular problem for \acp{hmd}, if one wants to benefit from the 3D rendering to break down spatial complexity in the medical data. They utilize high-resolution displays needing low-latency rendering to deliver immersive experiences and a smooth user experience, often without the benefit of being tethered to high-performance workstations. In many situations, like education or clinical settings, powerful workstations and tethered devices are difficult to utilize. Standalone immersive \acs{hmd} are affordable and easy to deploy and handle but severely constrained in storage and render performance, making it hard to maintain fast render speeds and high quality. 

To address these challenges, it is crucial to find a way to display medical volumetric data suitable for less powerful, mobile immersive devices. Rather than compromising rendering quality by opting for faster but less visually compelling techniques, lowering the resolution, or relying on aggressive denoising, we explore \ac{gs} as an intermediate representation for high-quality renderings of volumetric medical data. In our approach, the target anatomy or tissue is pre-rendered to high-quality images using volumetric path tracing, which in turn are used to optimize a \ac{gs} model. This representation can reproduce fine details, be compressed, and efficiently rendered in real-time on \ac{vr} systems, making it well-suited for mobile and immersive environments. Still, it has certain limitations. The \ac{gs} representation is inherently static, imposing several constraints: Color and transparency, commonly adjusted through transfer functions, are fixed at rendering time. Displaying different parts of the scanned anatomy would require multiple pre-trained, standalone \ac{gs} representations. Techniques like cutting, slicing, and isolating regions of interest are technically possible but produce visually incorrect results. The \ac{gs} optimization process primarily approximates surfaces and externally visible regions, with the dense inner parts of the volume largely missing. Mixing \ac{gs} representations can lead to visual artifacts, as the different volumes have not been optimized together and need to be globally sorted.

\begin{figure*}[ht!]
    \centering
    \includegraphics[width=\textwidth]{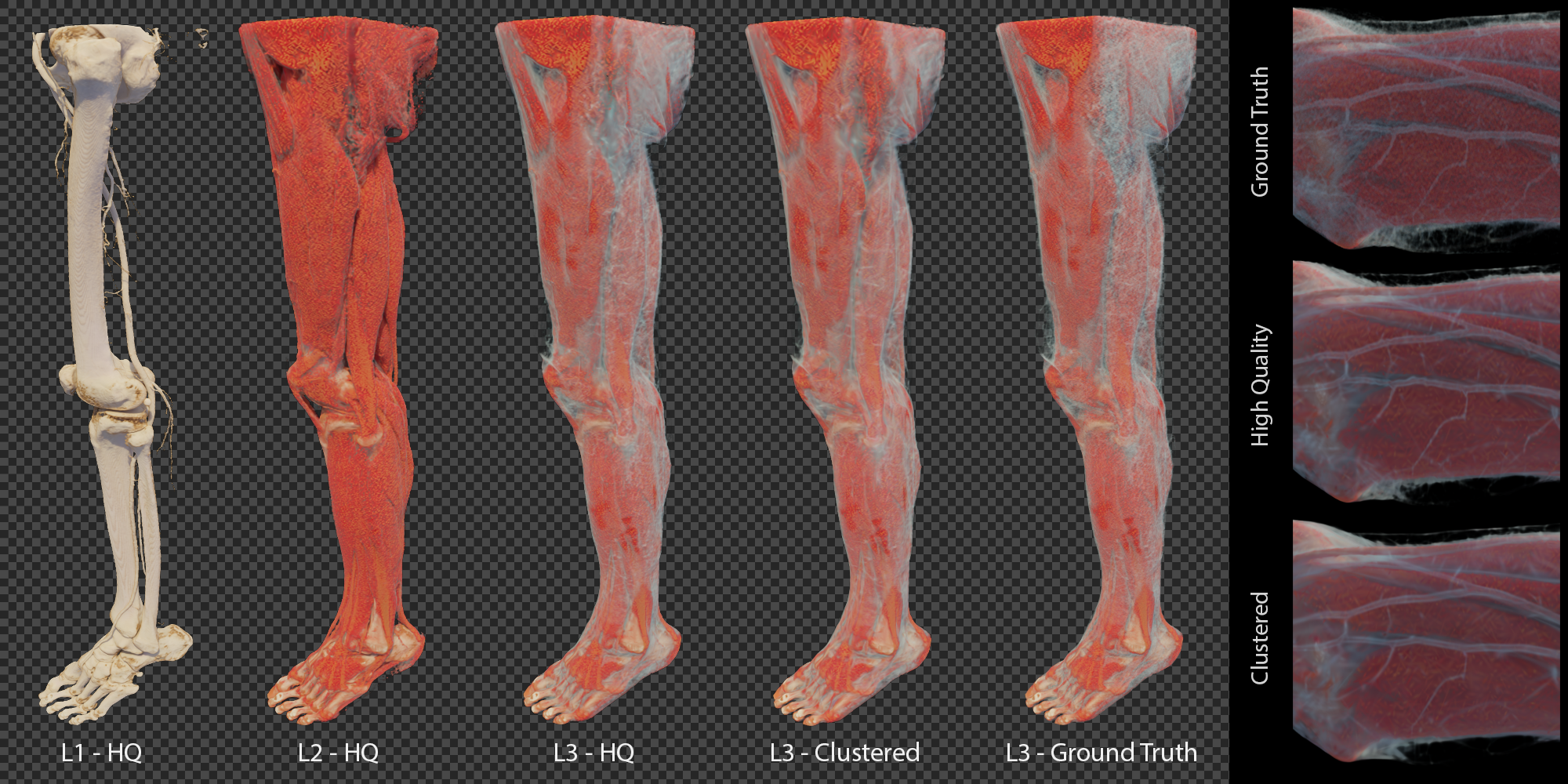}
    \caption{Visualization of the three layers in the leg scene from the high quality reconstruction, rendered with Unity. Comparison to default reconstruction with clustering compression and ground truth. Small details are lost or blurred in the reconstructions. }
    \label{fig:leg}
\end{figure*}

We propose a layered \ac{gs} representation to overcome these issues. Our Gaussian representation consists of multiple layers consecutively trained on top of each other, mitigating limitations associated with the static nature of \ac{gs}. With a layered representation, we can display anatomy with multiple embedded visualizations for the same structure. For example, a leg can be visualized with only the bone, bone and muscle, or bone, muscle, and soft tissue in one representation, as shown in \autoref{fig:leg}. The upper layers, such as muscle and soft tissue, would consist of Gaussians encoding only the additional information, while Gaussians from lower layers remain present and untouched. This allows for seamless switching and cutting between layers, similar to what is shown in \autoref{fig:teaser}, and the conditional display of additional information like highlighted areas with variable opacity. Our layered representation is more compact and faster to render than traditional path tracing of a \ac{ct} volume. It offers more dynamism and flexibility than normal \ac{gs} and can be more compact than storing multiple single representations with overlapping content.

\textbf{Contribution.}
Our work makes the following contributions:
\setlist{nolistsep}
\begin{enumerate}%
\item Insights into anatomy visualization using \ac{gs} on mobile and immersive display devices, with performance indications.%
\item A method for training and layering multiple \ac{gs} renderings on top of each other, storing them in a single splat asset.%
\item A technique for rendering these layered \ac{gs} representations in the Unity game engine, with the ability to cut and display them individually or in combination.%
\item Openly available training \& rendering code and datasets.
\end{enumerate}

\section{Related Work}

This section explores key techniques in medical volume visualization, discussing the evolution from \ac{dvr} to path tracing. We then examine recent advancements in \ac{nvs} techniques, particularly \ac{gs}, highlighting their applications and limitations in medical imaging contexts, and introducing alternative incremental training approaches.

\subsection{Volumetric Rendering}

Among various approaches to render volumetric medical data, \ac{dvr} has emerged as the most widely applied algorithm. It allows to render volumes as shaded 3D images without the need for intermediate representations~\cite{zhang_volume_2011}. \Ac{dvr} typically involves sampling along rays through the volume, filtering, and mapping volume density to RGBA values. These values are then blended to create the final image~\cite{engel_real-time_2006}. Colorization of the volume data usually occurs through the use of transfer functions. While more complex multi-dimensional transfer functions exist, most at least map from a sampled volume density to an RGBA color~\cite{ljung_state_2016}. Path tracing is an evolution of this rendering approach. It simulates physically based light transport by following the path of light around the scene after scattering events. This technique has been successfully applied to medical data, resulting in highly realistic and finely detailed imagery ~\cite{kroes_exposure_2012, dappa_cinematic_2016, comaniciu_shaping_2016}. Clinicians have rated path-traced renderings better than traditional DVR renderings~\cite{cardobi_path_2023}, although improved diagnostic value is not always clear~\cite{rowe_application_2019}. While there is interest in using volume rendering with immersive displays~\cite{kleinbeck_adaptive_2023, wheeler_virtual_2018, preim_survey_2018}, the high computational requirements, especially for path tracing, impose constraints on image quality. A common approach to accelerate rendering is the use of denoising techniques~\cite{huo_survey_2021}. Real-time capable approaches for medical data visualization exist~\cite{iglesias-guitian_real-time_2022, taibo_immersive_2024}, alongside offline methods targeting higher quality~\cite{hofmann_neural_2020}. Real-time methods typically require powerful hardware and often compromise on quality to maintain performance. Notably, none of these approaches deliver acceptable quality on mobile immersive devices. The benefits of using path-traced volume visualizations include more realistic rendering~\cite{glemser_new_2018}, improved surface detail~\cite{rowe_application_2019}, and enhanced perception of complex anatomic spatial relationships~\cite{caton_jr_three-dimensional_2020}, making path tracing a valuable tool in medical visualization, despite its computational challenges.

\subsection{Novel View Synthesis}

\Ac{nvs} refers to the generation of new viewpoints of a scene from a set of existing images or viewpoints. Pioneered around 30 years ago, it has evolved from techniques like light field rendering and view interpolation~\cite{levoy_light_1996}. Today, most \ac{nvs} techniques involve \acp{nn} in some form. \Ac{nerf} have brought a leap in quality and realism to \ac{nvs}~\cite{mildenhall_nerf_2020}, however, at the cost of long training times, slow rendering, and extensive storage requirements. Numerous approaches inspired by or derived from \ac{nerf} have emerged, many aiming to optimize its performance shortcomings~\cite{muller_instant_2022, li_compressing_2022}. Despite these efforts, there is no silver bullet for addressing all limitations simultaneously, which makes immersive applications particularly challenging due to the massive performance requirements~\cite{li_rt-nerf_2022, deng_fov-nerf_2022, rolff_interactive_2023}. Falling back to mesh exports to leverage existing rendering pipelines can bring neural reconstructions to immersive mobile displays, however, with losses in image quality~\cite{kleinbeck_neural_2024}.

\subsubsection{Gaussian Splatting} 
Kerbl et al.~\cite{kerbl3Dgaussians} introduced an explicit Gaussian representation for radiance fields, known as \ac{gs}. It is a variant of \ac{nvs} initialized from a sparse point cloud, followed by optimization of the Gaussians. The final image is then created by splatting these optimized Gaussians~\cite{zwicker_ewa_2001}. 
Unlike \acs{nerf}, \ac{gs} does not require a \ac{nn} and uses efficient rasterization to render the optimized scene representation.
This generally results in faster runtime performance compared to \ac{nerf}, making it a promising candidate for \ac{vr} applications~\cite{jiang_vr-gs_2024}. Yet, the original implementation also comes with substantial storage requirements, which has triggered the development of various compression approaches~\cite{bagdasarian_3dgszip_2024, morgenstern_compact_2024}. Some of these approaches have also improved rendering speed~\cite{navaneet_compact3d_2024} and fidelity~\cite{chen_hac_2024}. \Ac{gs} has been extended with anti-aliasing through a low-pass and 2D Mip filter, improving visual quality at focal lengths and rendering distances different from the original training data~\cite{yu_mip-splatting_2023}. Similar to our layering approach exist works that store Gaussians in somewhat independent chunks, enabling smooth transitions between them~\cite{kerbl_hierarchical_2024}. They do target large-scale scenes by employing a hierarchical structure, though different from our overlapping data. Given the approach's novelty, further speed and quality improvements will likely emerge.

In the medical field, many \ac{gs} approaches focus on minimally endoscopic images with deformable structures. Some represent the structures and changes over time using a \ac{mlp}~\cite{liu2024lgslightweight4dGaussian,xie2024surgicalGaussiandeformable3dGaussians,zhao2024hfgs4dGaussiansplatting}. Utilizing depth generated from a foundation model~\cite{yang_depth_2024} can be useful to optimize \ac{gs} for similar structures in endoscopic images~\cite{liu_endogaussian_2024,zhao2024hfgs4dGaussiansplatting}. These approaches focus on the deformability of the Gaussians due to the anatomical structure of organs such as the gallbladder. Another recent work has explored combining cinematic rendering~\cite{dappa_cinematic_2016}, a form of path tracing, with \ac{gs} for anatomy visualization~\cite{niedermayr_application_2024}. They integrate view selection algorithms, alpha channel training, and anti-aliasing for quality improvements. The study evaluated \ac{gs} using more powerful hardware focusing on highly detailed, static representations, finding the approach suitable for anatomy visualization.

\subsubsection{Continual Learning for Novel View Synthesis}

In \ac{nvs}, works exploring consecutive training approaches usually fall within the domain of continual learning. One challenge in continual learning is catastrophic forgetting~\cite{mccloskey_catastrophic_1989}. To prevent this, the parameters are isolated in updating learning step~\cite{serra_overcoming_2018, mallya_packnet_2018}. An approach in \ac{nerf} uses chamber loss instead of the classical photometric loss~\cite{chung_meil-nerf_2022}. Other approaches query a pre-trained \ac{ngp} for pseudo ground truth and train a new \ac{mlp} for the current scene~\cite{po_instant_2023}, or use a replay buffer to address this issue~\cite{cai_clnerf_2023}. While our approach shares similarities with existing continual learning \ac{nvs} approaches, our main goal is to maintain distinct scenes by freezing previous layers. We aim for a full layered representation of anatomy instead of the same scene with various differences over time or a consecutive capturing setup.

\begin{figure*}[t!]
    \centering
    \includegraphics[width=\linewidth]{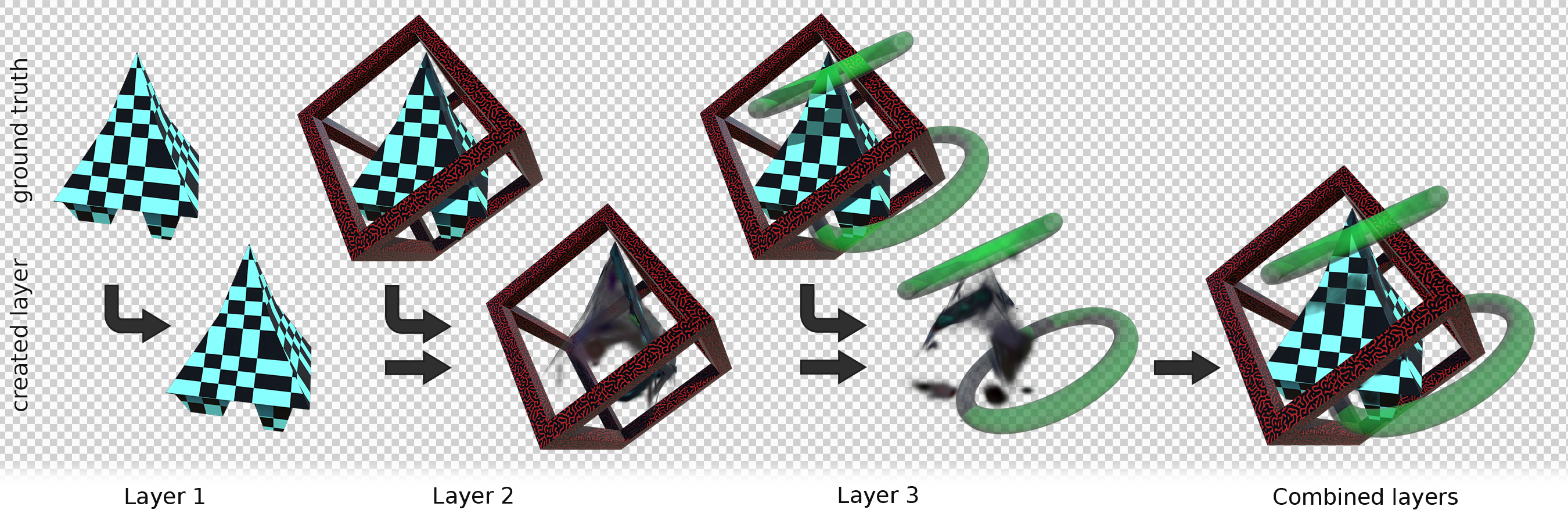}
    \caption{Synthetic three layer scene. Top row show the three layers of ground truth data. Bottom row shows the active layer trained from the corresponding images, without previous layers, and a combined view. Each layer consists predominantly of newly added elements and areas where existing elements are shadowed.}
    \label{fig:3layer}
\end{figure*}

\subsection{Summary}

Recognizing the limitations of static \ac{gs} representations, particularly their inability to adjust in real-time, we explore the training of multiple visualizations of the same medical volume into a single \ac{gs} representation. The original \ac{gs} training process was primarily designed for natural scenes without transparent backgrounds. Although it supports synthetic data with transparency, the alpha channel is flattened during loading. This behavior can lead to artifacts and loss of detail in semi-transparent areas in front of transparent backgrounds, as demonstrated in \autoref{fig:transparency}. 

\section{Method}

To bring high-quality and realistic anatomical visualizations to moderately powered hardware, such as mobile \ac{vr} headsets, our approach utilizes precomputed high-quality path-traced images to train a \ac{gs} intermediate representation. Specifically, we propose a layered \ac{gs} model, where lower layers act as a "background" for subsequent layers, creating a composite representation, as shown in \autoref{fig:teaser}. Our anatomy rendering approach is structured into three distinct steps, each of which can occur at different points in time:
\begin{enumerate*}[label=(\roman*)]
\item image generation with a path tracer,
\item train the layered \ac{gs} representation using the rendered images,
\item render the trained \ac{gs} anatomy representation on the target device.
\end{enumerate*}
These three steps are described in more detail in the following sections.

\subsection{Image generation}
\label{sec:imagegeneration}
For this experiment, we adapted an open-source path tracer\footnote{\url{https://github.com/nihofm/volren}}. For image generation, we scripted the path tracer to capture images from multiple angles around the object of interest. The renderer employs Monte Carlo volume path tracing. Color and density are accumulated along the traced paths in the medical volume via RGBA values defined by a transfer function. Scattering events are modeled using a Henyey-Greenstein phase function, and lighting contributions are sampled from a high dynamic range environment light map. The final rendered images are tone-mapped before saving to ensure proper dynamic range and brightness for downstream applications. We note that some features found in other path tracers, such as specularity, can be approximated by \ac{gs} through spherical harmonics. In contrast, others, like depth of field, cannot be encoded and should be disabled.

\Ac{gs} starts optimizing from an initial (sparse) point cloud. Typically, this would be either a random point cloud for synthetic images or generated by a structure-from-motion approach like COLMAP~\cite{schonberger_structure--motion_2016} for real images. Since we have full control over the volumetric data used to render the training images, we directly generate the initial point cloud. We place points where the approximate surface of the volume will be to ensure an effective basis for Gaussian color and position. Directly translating volume voxels into a point cloud would result in most points being located inside the volume rather than on its surface. Instead, we sample points from outside the volume inward, stopping and creating a point as soon as a minimum transfer function opacity $alpha\_treshold = 0.2$ is reached. This approach results in a surface approximation of the volume close to where the final Gaussians will be located, with correct point color. We export this point cloud, along with the created images and their corresponding camera positions and descriptions, in COLMAP format to enable the direct use in \ac{gs} training. 

\subsection{Layered Training}

We extend conventional \ac{gs} with a layering system, enabling the inclusion of multiple distinct visual representations within a single \ac{gs} point cloud. Our layering technique is particularly useful for anatomical visualization, where, for instance, layers can be trained sequentially to represent bone, muscle, and soft tissue. This is achieved through consecutive training iterations utilizing the previously optimized \ac{gs} as a base. In our approach, the previously optimized \ac{gs} layers remain static, functioning similarly to frozen layers in \acp{nn}~\cite{serra_overcoming_2018, mallya_packnet_2018}. During the training of each new layer, the existing layers are excluded from adjustments by the optimizer, ensuring they maintain their visual integrity even in the absence of the newly added layer.

The pre-optimized point clouds serve as an advanced 3D background in the rendering process, contributing to the image loss. All layers are finally consolidated into a single \ac{gs} .ply file, with layer membership information stored as a property alongside attributes like color and scale. As progressive layers are optimized with previous layers present, the optimization process primarily adjusts the Gaussians to represent new information, while existing lower layers contribute visual information when possible. Consequently, each new layer effectively contains a differential representation, encoding only the new data not captured by earlier layers. 

\subsubsection{Alpha Channel Training}
In addition to layering, we extended the training process to preserve and utilize the alpha channel of input data. We employ two strategies: randomizing the background color for each frame instead of flattening it at the start and incorporating the alpha channel into the loss calculation. Randomized background colors improve detail in semi-transparent areas, while including the alpha channel ensures that the number of Gaussians needed to achieve this detail remains manageable. 

\begin{figure}[t!]
    \centering
    \includegraphics[width=\columnwidth]{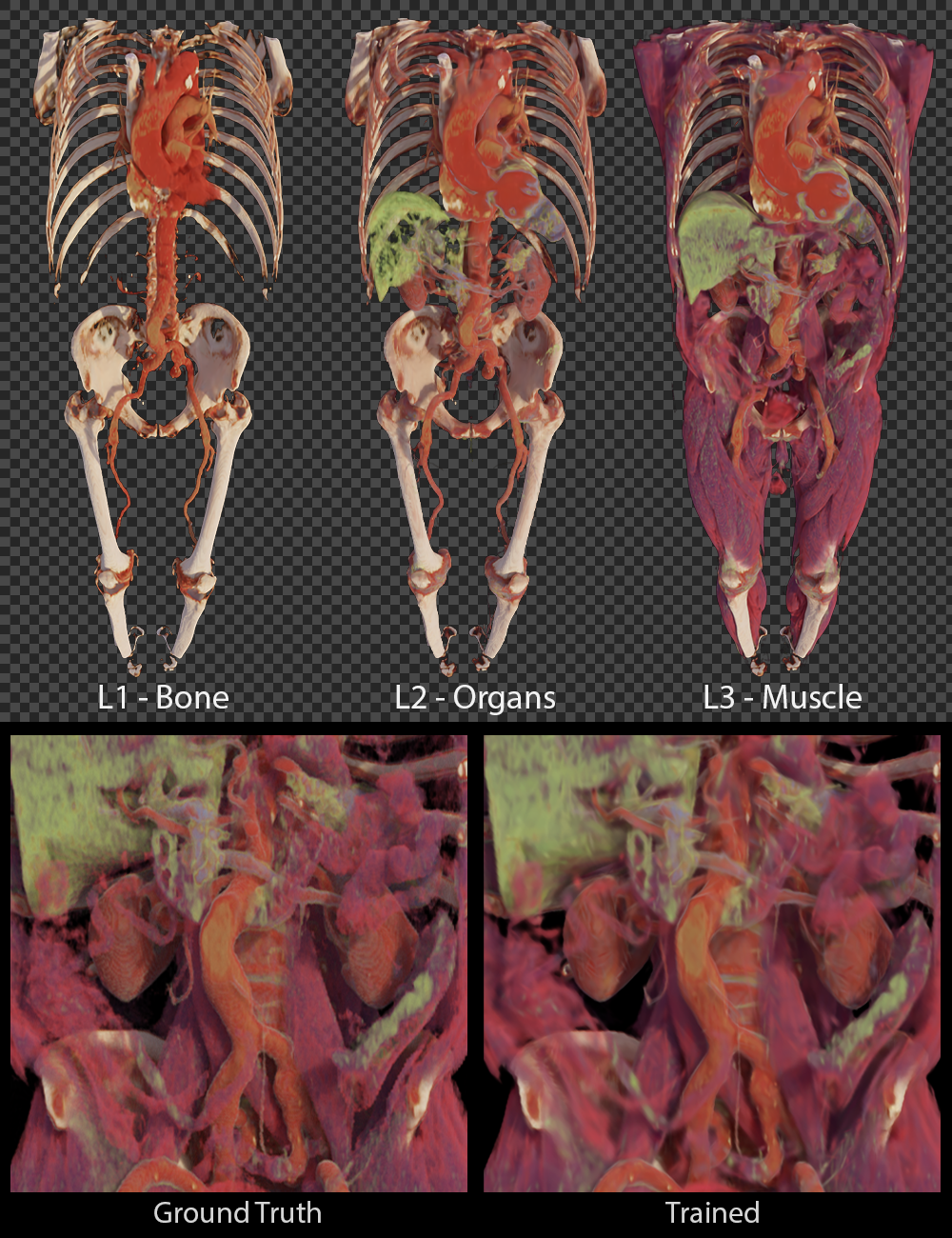}
    \caption{Fullbody CT scene consisting of the three layers Bone, Organs and Muscle. Rendered with Unity and clustering compression applied. Structures are largely present, but colors are smoothed.}
    \label{fig:fullbody}
\end{figure}

\subsubsection{Inactive Pruning}
The standard \ac{gs} optimization process also does not account for the complexities introduced by having "frozen" layers of Gaussians that contribute to the final rendered image but cannot be adjusted further. This is particularly problematic when dealing with Gaussians that do not contribute to the scene but are masked by existing layers. We extend the \ac{gs} adaptive control with an additional pruning step to remove inactive Gaussians from the scene. We track the optimization of each Gaussian (position, color, scale) and remove inactive Gaussians that fail to meet a minimal activity threshold. These Gaussians are assumed to be either obscured by already optimized layers or already very well optimized, which is unlikely to occur early in the training. Gaussian activity is tracked in sync with the existing densification rate and compared against a threshold. After each pruning step, we apply exponential decay.

The average activity is computed as follows, with a densification rate \( dR = 100 \), position norm \( p \), color norm \( c \), and scale norm \( s \):
\[
\text{AvgSample} = \frac{1}{dR} \sum_{i=1}^{dR} \left(2 \cdot p_i + 0.1 \cdot c_i + 0.1 \cdot s_i\right)
\]
This average is compared against an initial threshold \( T = 0.015 \):
\[
\text{AvgSample} \leq T
\]
After each comparison, the threshold is updated using exponential decay:
\[
T_{\text{new}} = 0.975 \times T_{\text{old}}
\]
The three activity components are weighted to have a comparable influence based on our data. The initial threshold and decay rate values were experimentally determined using our datasets. We tested a range of values to strike a balance between the number of removed Gaussians and the impact on visual quality. Our approach filters out low-activity Gaussians early on, tapering off to avoid accidental removal of well-optimized Gaussians contributing to the scene. Although our default values were effective for a range of scenes in our tests, they can be fine-tuned for specific scenes to achieve optimal results.

\begin{figure}[t!]
    \centering
    \includegraphics[width=\columnwidth]{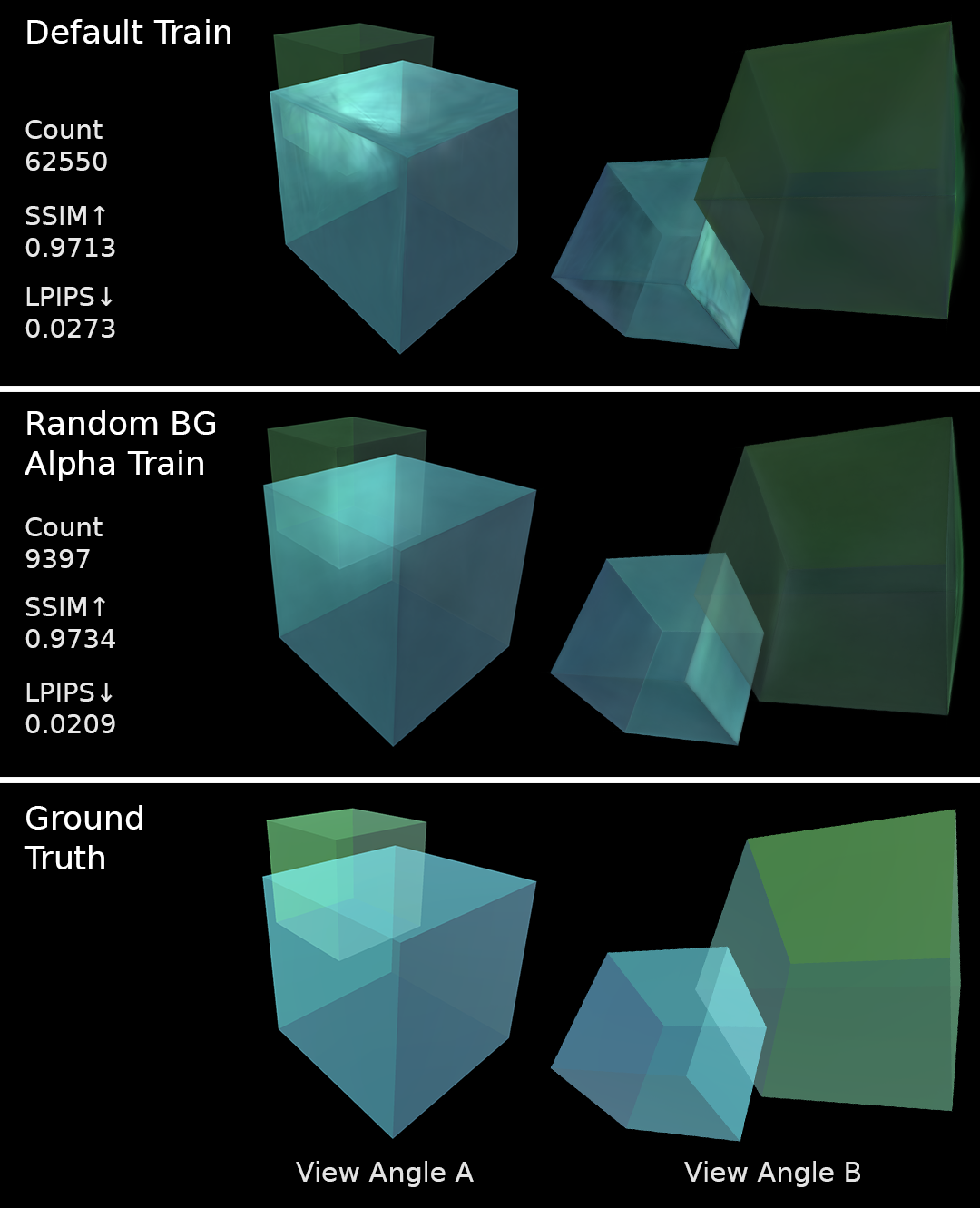}
    \caption{Two viewpoints of synthetic transparent data trained without (first row) and with (second row) alpha channel training and randomized backgrounds. Both help to reduce visual artifacts while lowering the Gaussian count. Due to different alpha blending between applications, \ac{gs} can struggle to faithfully recreate translucent regions.}
    \label{fig:transparency}
\end{figure}

\subsection{Real-time Rendering}

\begin{table*}[]
\centering
\caption{Comparison of sizes between layered and stand-alone training. Standalone does generally incur higher storage requirements compared to layered Gaussians. The reported size are before and after clustering compression, plus the number of Gaussians in parenthesis. The size and Gaussian count for layered representations include all lower layers as they are required for accurate rendering.}
\begin{tabularx}{\linewidth}{@{}lYYYYYY@{}}
\toprule
         & \multicolumn{2}{c}{Leg} & \multicolumn{2}{c}{Fullbody} & \multicolumn{2}{c}{Synthetic 3Layer} \\ \cmidrule(l){2-3} \cmidrule(l){4-5} \cmidrule(l){6-7}
         & \scriptsize Layered &  \scriptsize Standalone & \scriptsize Layered & \scriptsize Standalone  & \scriptsize Layered & \scriptsize Standalone      \\ \midrule
Layer 1  & 14.5/3.4 (57k)      & 14.5/3.4 (57k)          & 25.0/4.7 (99k)      & 25.0/4.7 (99k)          & 17.3/3.7 (69k)  & 17.3/3.7 (69k)  \\
Layer 2  & 30.8/5.4 (122k)     & 17.2/3.8 (68k)          & 34.7/5.9 (138k)     & 20.4/4.2 (81k)          & 40.0/6.5 (159k) & 29.9/5.2 (119k) \\
Layer 3  & 34.4/5.8 (137k)     & 14.9/3.5 (59k)          & 49.3/7.6 (196k)     & 20.9/4.2 (83k)          & 41.7/6.7 (166k) & 25.6/4.7 (102k) \\ \midrule
Combined & 34.4/5.8 (1379k)    & 46.6/10.7 (184k)        & 49.3/7.6 (196k)     & 66.3/13.1 (262k)        & 41.7/6.7 (166k) & 72.8/13.6 (291k)\\ \bottomrule 
\end{tabularx}
\label{tab:layeredsizecomparison}
\end{table*}

We utilize the Unity game engine in version 2022.3 as our platform for rendering due to its broad compatibility, including most \ac{vr} headsets. To display our trained layered \ac{gs} representation, we extend an open-source \ac{gs} rendering package for Unity\footnote{\url{https://github.com/aras-p/UnityGaussianSplatting}}. The rendering process is analogous to that described by Kerbl et al.~\cite{kerbl3Dgaussians}, but implemented platform-agnostically in HLSL and compute shaders. The rendering employs wave operations to achieve real-time rendering, making it incompatible with older devices and OpenGL rendering, but should otherwise export to most modern platforms. Notably, the implementation does not rely on CUDA. The rendering of our \ac{gs} representations involves two main steps. First, the \ac{gs} point cloud file is imported, compressed, and transformed into GPU-friendly buffers for rapid access. At runtime, the buffers are transferred to the device's GPU. The Gaussians are continuously sorted relative to the virtual camera position and rendered as splats. To optimize resource usage, we only  transfer actively rendered layers to the GPU, while unused layers are kept in RAM.

While compression is not the primary focus of this work, it is crucial for both reducing storage footprint and GPU memory consumption, especially on mobile devices. Through the Unity rendering package, we apply quantization to the Gaussian parameters. Since spherical harmonics coefficients are a significant portion of the file size (approximately $3/4$ of the total size) and are the least noticeable when modified, we use K-means clustering to further compress them, similar to~\cite{navaneet_compact3d_2024}. This is particularly effective because we cluster across all input file layers, allowing information to be shared. Volume renderings typically use a limited color palette, further enhancing the effectiveness of this approach. An example of our Unity rendering can be seen in \autoref{fig:fullbody}.

The Gaussians must be constantly sorted relative to the viewer during rendering to ensure correct display. In \ac{vr}, this sorting must be performed twice—once for each eye. The sorting of splats can differ slightly between the two eyes; however, it does mostly for splats very close to the rendering camera. We explore an optimization in which the sorting is shared between both eyes, effectively halving the sorting time per frame. We use two sorting algorithms: AMD FidelityFX Parallel Sort (FFXPS) and an implementation derived from OneSweep~\cite{adinets_onesweep_2022}, DeviceRadixSort. DeviceRadixSort is roughly twice as fast as FFXPS but faces compatibility issues with many mobile devices through Unity shader transpilation.

\section{Evaluation}

Our approach enables the integration of multiple \ac{gs} representations into a single cohesive scene by overlaying anatomical layers, rather than having isolated representations. In this section, we analyze the storage and computational requirements of our method, along with the impacts of extensions such as inactive pruning and alpha channel training. We evaluate our approach using both real-world CT scans and specifically created synthetic data.

\subsection{Metrics}

To compare the rendering quality of the novel views, we report \ac{psnr}, \ac{ssim}~\cite{wang_image_2004} and \ac{lpips}~\cite{zhang_unreasonable_2018}. We report file sizes in \textit{MB}. Where possible, values are reported with one standard deviation after ± sign.

\subsection{Datasets}

To test our layered approach, we generate two datasets: one from real \ac{ct} scans and another with synthetic images. The anatomies we use are \textbf{Lower Body}~\cite{tcga-hnsc}\footnote{TCGA-HNSC dataset, id \texttt{TCGA-CV-A6JU}, \url{https://www.cancerimagingarchive.net/collection/tcga-hnsc/}}, a human lower body \ac{ct} scan. It has a high resolution of $512 \times 512 \times 3000$ voxels at $0.7 \times 0.7 \times 0.3$ mm each. The \textbf{Legs} scan is one leg isolated from the lower body scan. It has a resolution of $195\times 257 \times 3000$ voxels at unchanged dimensions. The \textbf{Fullbody} scan\footnote{TotalSegmentator dataset~\cite{wasserthal_totalsegmentator_2023}, id \texttt{s0287}, \url{https://zenodo.org/records/10047292}} is a near-complete human body scan from neck to ankles. We removed the patient bed and cropped the scan to $317 \times 163 \times 835$ voxels. Each anatomy scene includes 512 images, of which we use every 8th for testing, each 1600 px $\times$ 1600 px, from various poses.

Additionally, we created synthetic scenes to highlight layer reconstruction and investigate potential problematic cases. Each comprises 256 training and 32 test images at 1400 px$ \times $1400 px. The \textbf{Two Layer} scene contains a plain white cube in the first layer; the outer layer is a 5\% bigger cube with a white and blue pattern. The scene allows testing \ac{gs} approaches for redundant Gaussians or ones located in obscured areas. The \textbf{Three Layer} scene assesses the definition of each layer in scenarios with sparse overlap and complex shapes. The first layer is an opaque green and black checkered shape, the second layer is a red and black cube outline around the first layer, and the third layer includes two semi-transparent green rings looping through the second layer. Aside from layering, the \textbf{Transparency} scene is a single-layer scene featuring two cubes with varying levels of transparency. The light green cube is 75\% transparent, while the light blue cube is 62.5\% transparent.

\begin{table}[t]
\centering
\caption{Comparison of medical volumes on-disk sizes before and after training and compression. The size of the reconstruction is independent of the original volume resolution and more influenced by the visual complexity of the rendered anatomy.} \label{tab:sizecomparison}
    \begin{tabularx}{\linewidth}{@{}lYYY@{}}\toprule
                                      & Leg     & Lowerbody  & Fullbody \\ \midrule
        Original Volume (DICOM)       & 304.7   & 1585.1     & 87.4     \\
        Compressed (.nii.gz)          & 73.5    & 730.8      & 75.5     \\ \midrule
        \ac{gs} PLY file (ours)       & 34.4    & 23.0       & 49.3     \\
        Layered GS (default comp.)    & 12.8    & 8.6        & 18.2     \\
        Layered GS (clustering)       & 5.8     & 4.5        & 7.6      \\ \bottomrule
    \end{tabularx}
\label{tab:filesizecomparison}
\end{table}

\subsection{Parametrization Details}

Optimizing our \ac{gs} representations, we balanced image quality and rendering speed to achieve real-time performance on immersive displays. These targets are directly influenced by the number of Gaussians generated during the optimization process, which is, in turn, affected by the training parameters. We train two representations of our anatomic data to gauge the impact of parameter choice and Gaussian count. The default reconstruction uses the default \ac{gs} parameters; for the training of the high quality (\emph{HQ}) one, we set $densify\_grad\_threshold = 0.0001$ and $percent\_dense = 0.005$, resulting in a roughly two to fourfold increase in the number of Gaussians created. All anatomy scenes use initial point clouds with approximately 100,000 points.

We further evaluate two compression levels. We either use \emph{low compression}, converting point cloud parameters to 16-bit floats and reducing spherical harmonics coefficients to 6 bytes per triplet, roughly 10.6 bits per color channel. For high compression, we apply more aggressive quantization and use K-means clustering to compress spherical harmonics coefficients into 16k known variants. This compression level is referred to as \emph{clustered} throughout the work. Comparisons of file sizes can be found in \autoref{tab:sizecomparison}.

\begin{table*}[t!]
    \centering
    \caption{Rendering time in ms \& quality comparison on the leg scene. Images were rendered at 2000 px $\times$ 2000 px $\times$ 2 (exemplary \ac{vr} device) on a PC with an Nvidia RTX 4090 GPU. Path traces with 4 bounces. Values are reported over all test images with one standard deviation. DVR values are illustrative examples, as the rendered images need to be brightness adjusted and will always be slightly different. Denoising with Open Image Denoise, raw render time parenthesis.}
    \begin{tabularx}{\linewidth}{lccYcYc} \toprule
                &  \multicolumn{3}{c}{Path tracer} &  \multicolumn{3}{c}{Layered GS in Unity (Ours)} \\  \cmidrule(l){2-4} \cmidrule(l){5-7}
                & \scriptsize 1024 samples & \scriptsize 1 sample + denoise & \scriptsize \ac{dvr} & \scriptsize HQ recon/low comp. & \scriptsize low comp. &  \scriptsize high compression \\ \midrule
Time (ms)        & 7116.4 ± 2251.3   &  66.0 ± 10.6 (19.9)&  17.9 ± 10.2      &  1.7 ± 0.4        &  1.3 ± 0.4       & 1.4 ± 0.4        \\
PSNR~~↑           & 42.399 ± 2.198    &  26.658 ± 2.596    &  24.258 ± 3.514   &  37.662 ± 2.335   &  35.463 ± 3.170  & 35.258 ± 3.088   \\
SSIM~~↑           & 0.988 ± 0.005     &  0.945 ± 0.018     &  0.914 ± 0.033    &  0.969 ± 0.012    &  0.964 ± 0.014   & 0.963 ± 0.015    \\
LPIPS~↓          & 0.020 ± 0.010     &  0.081 ± 0.032     &  0.083 ± 0.033    &  0.054 ± 0.025    &  0.062 ± 0.029   & 0.063 ± 0.029    \\ \bottomrule

    \end{tabularx}
    \label{tab:rendertimequality}
\end{table*}

\begin{table}[t]
\centering
\caption{Rendering time of Leg scene in ms on different \ac{hmd} devices at their native resolution. Varjo Aero is a desktop \ac{vr} system driven by an Nvidia RTX 4090, Meta Quest 3 is a standalone mobile \ac{vr} system, MacBook is an Apple laptop with M3 Pro CPU/GPU.} \label{tab:render_vr}
    \begin{tabularx}{\columnwidth}{@{}ll|YYY@{}}\toprule
        Sort     & HQ        & Varjo Aero   & Meta Quest 3  & MacBook      \\ \midrule
        \cmark   & \xmark    & 2.6 ± 0.8    & 17.6 ± 5.9    & 4.8 ± 1.5    \\
        \xmark   & \xmark    & 2.5 ± 0.8    & 10.2 ± 2.1    & 4.5 ± 1.3    \\
        \cmark   & \cmark    & 3.6 ± 1.0    & 26.6 ± 7.0    & 6.4 ± 1.9    \\ 
        \xmark   & \cmark    & 3.4 ± 0.9    & 16.3 ± 2.2    &  6.0 ± 1.6   \\ \bottomrule
    \end{tabularx}
\label{tab:devicespeedcomparison}
\end{table}

\begin{figure}[t]
    \centering
    \includegraphics[width=1.0\linewidth]{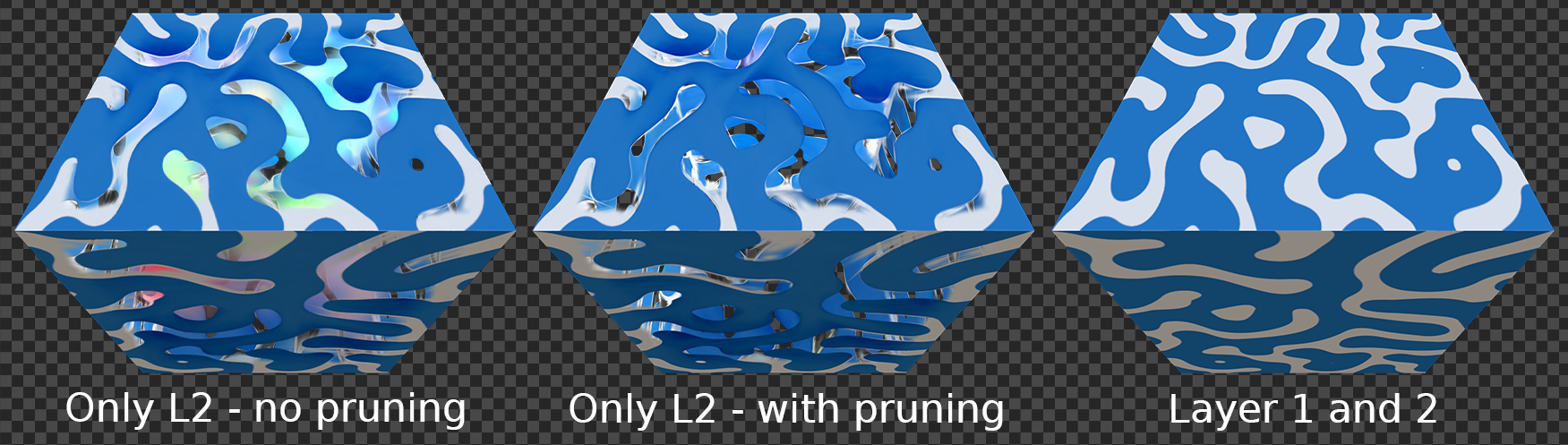}
    \caption{Synthetic two-layer scene, trained without and with inactive pruning. The left and middle segments show only the second layer which does not contain the white pattern, as the lower layer provides them. Notice the colored artifacts of unoptimized Gaussians normally covered by the lower layer, removed by inactive pruning.}
    \label{fig:2layer-prune}
\end{figure}

\subsection{Performance Analysis}

\subsubsection{Speed}
As shown in \autoref{tab:rendertimequality}, our \ac{gs} representation outperforms classical and denoised path tracing, as well as \ac{dvr}. This performance advantage is notable even though our path tracer uses advanced techniques, such as a multi-level digital differential analyzer, to improve speed and reduce memory usage~\cite{hofmann_efficient_2021}. It is also important to consider pathological cases: In point cloud rendering, performance is more predictable and depends largely on the Gaussian count. In contrast, volume rendering speed is tied to the number of pixels requiring color calculations, leading to performance breakdowns during close-up views. In worst-case scenarios in our test set, some images can exceed four times the average render time, compared to around 1.6 times for \ac{gs}. Rendering on mobile and desktop VR devices performs well overall. However, standalone immersive devices like the Quest 3 show performance bottlenecks, as seen in~\autoref{tab:devicespeedcomparison}. We found that sorting Gaussians for the eye center only, rather than both eyes, and reducing sorting frequency can enhance frame rates with minor visual compromises. These findings suggest a potential for high-quality anatomical visualization on hardware with limited capabilities. Note that the Unity \ac{gs} plugin we built upon has limited compatibility across mobile headset platforms and Unity render pipelines. On platforms supporting both the built-in and URP render pipelines, we observed no significant performance differences.

\subsubsection{Quality}
\Ac{gs} effectively reconstructs complex anatomies from path-traced volumetric inputs in our test scenes, achieving high visual quality scores~(see \autoref{tab:rendertimequality}). Each layer can be rendered independently but should be displayed with all lower layers for optimal results as trained. While internal views of trained \ac{gs} models may appear messy, frontal and side views can provide interactive, engaging visualizations of otherwise static representations (see \autoref{fig:teaser}). Reconstruction quality primarily depends on the number of Gaussians, generally outperforming path tracing at low sample sizes (see \autoref{tab:rendertimequality}). However, fine details may appear blurry, especially in low-frequency regions. High-contrast features are accurately reconstructed when sufficient contrast is available. In close-ups, individual Gaussians are often visible. The quality of the initial point cloud also influences the optimization process. A well-structured initial point cloud leads to better early visuals and improved overall results. Random initialization may struggle to distribute Gaussians effectively in sparse areas, leading to smudging artifacts. We experimented with antialiasing as described in~\cite{yu_mip-splatting_2023}, but observed minimal improvements, likely due to our dataset's low variation in camera distances.

\subsubsection{Size}
\Ac{gs} offers significant storage advantages over traditional volume files for path tracing, typically stored in DICOM format (see \autoref{tab:filesizecomparison}). File sizes depend primarily on the applied transfer function and visual complexity of the anatomy, rather than volume resolution. Size reductions of up to 99\% compared to original DICOM files are achievable. While many small details can be preserved (compare \autoref{fig:fullbody}), this comes at the cost of spatial accuracy. Still, this reduction can benefit deployment to and storage on mobile devices. Although \ac{gs} parameters can be adjusted to control quality and size, fine-tuning this balance remains challenging.

\subsection{Layering}
Our layering approach integrates well with the GS optimization process. Gaussians are generally positioned where changes are most needed during optimization, keeping areas already well-represented by lower layers relatively free from unnecessary Gaussians (see \autoref{fig:3layer}). Layering is particularly efficient when there is minimal overlap between layers. The number of Gaussians required to encode new information depends on how much of the scene is obscured by the new layer and its visual complexity. Generally, this approach results in a smaller file size than training each target scene independently, as shown in \autoref{tab:layeredsizecomparison}. However, we also identified issues during optimization and subsequent rendering of upper layers, as the presence of background layers containing relevant information was not considered in the original paper. Our inactive pruning approach addresses these issues.

\subsection{Inactive Pruning}

When optimizing a layer with one or more already finished layers loaded, some areas might be wholly obscured from all camera views. Certain areas may become completely obscured from all camera views, such as the insides of bones. Gaussians inside these areas appear perfectly optimized during the optimization process, even though they add no information to the rendered scene. Since the optimization process does not recognize that these Gaussians are redundant, they are neither pruned nor relocated. Similarly, Gaussians that are close to already optimized layers and happen to look similar are also not pruned or moved, even though they contribute little new information. Our inactive pruning method removes these and other inactive Gaussians, as seen in \autoref{fig:2layer-prune}.

\subsection{Alpha Channel Training}

In the original \ac{gs} training procedure, the alpha channel was not considered. This omission is particularly problematic when dealing with synthetic renders and transparent backgrounds, as it can cause the background color to bleed into semi-transparent areas, causing artifacts. This effect is most pronounced with few views of the semi-transparent area. As shown in \autoref{fig:transparency}, training with randomized background colors and including the alpha channel in the loss function improves visual fidelity. Randomized background colors help most to improve visual quality, while including the alpha channel reduces the number of Gaussians needed to achieve this improvement.

\begin{figure}[t!]
    \centering
    \includegraphics[width=\columnwidth]{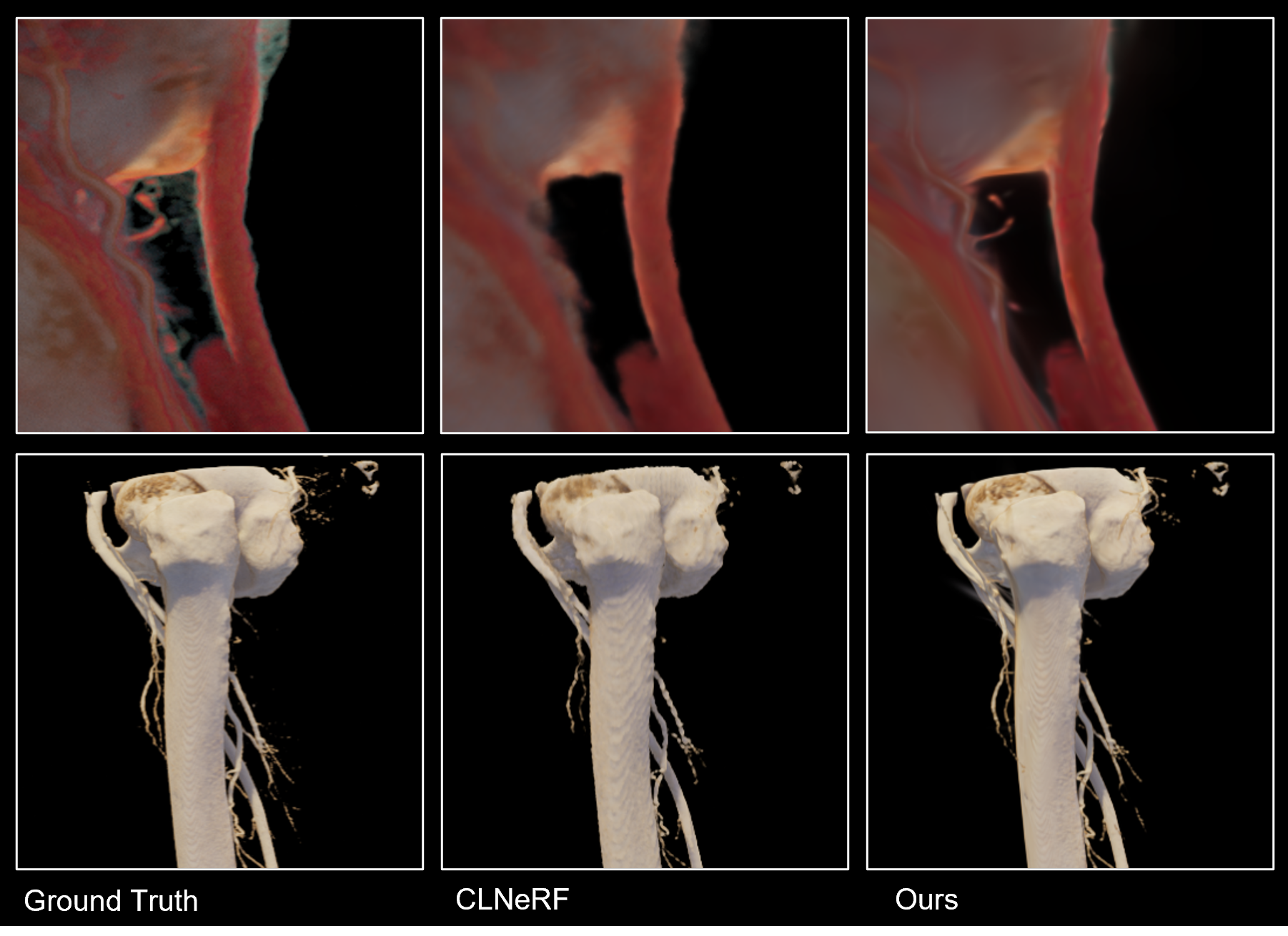}
    \caption{Comparison of ours and CLNeRF~\cite{cai_clnerf_2023} on the Leg scene.}
    \label{fig:clnerf_ours}
\end{figure}

\subsection{Continual Learning and Novel View Synthesis}

Since our work can be seen as a part of continual learning, we compared our work to CLNeRF~\cite{cai_clnerf_2023} on layered scenes of our synthetic dataset. CLNeRF and other continual learning approaches~\cite{chung_meil-nerf_2022,po_instant_2023} usually assume a stacking of the scene. Due to the high memory consumption of CLNeRF and high image resolution, we used a subset for comparing to continual learning, adapting the dataset using every 4th image per layer, starting from an incremental index to avoid exactly matching camera poses, which is the commonly used data structure in CLNeRF. 

Instant-NGP-based CLNerf is challenged by fine structures of anatomical data and bones show ripple artifacts. Although the results in \autoref{tab:cont_learn} show promising results, our approach can better represent the fine structures, even with fewer images. Moreover, the rendering time increases due to the replay buffer prevent real-time \ac{vr} use. To compare anatomic data, we use images with a resolution of $1600$x$1600$, following their original implementation. 

\begin{table}[t!]
    \centering
     \caption{Continual Learning comparison on reduced Leg \& Fullbody scenes.}

    \addtolength{\tabcolsep}{-0.4em}
    \begin{tabularx}{\linewidth}{@{}lYYYYYY@{}} \toprule
                  & \multicolumn{3}{c}{CLNeRF~\cite{cai_clnerf_2023}}  & \multicolumn{3}{c}{Ours}  \\ \cmidrule(l){2-4} \cmidrule(l){5-7}
                  & \scalebox{0.78}{ PSNR~↑} & \scalebox{0.78}{SSIM~↑} & \scalebox{0.78}{ LPIPS~↓} & \scalebox{0.78}{ PSNR~↑} & \scalebox{0.78}{ SSIM~↑} & \scalebox{0.78}{ LPIPS~↓} \\  \midrule
        Leg       & \textbf{39.952} & \textbf{0.974} & 0.042 & 35.481 & \textbf{0.974} & \textbf{0.040}\\
        Fullbody & 35.850 & 0.964 & 0.064 & \textbf{37.378} & \textbf{0.973} & \textbf{0.041}\\ \bottomrule
    \end{tabularx}
   
    \label{tab:cont_learn}
\end{table}

\section{Discussion}
In this work, we demonstrate that \ac{gs} can create useful representations of path-traced volumetric anatomy data. While a careful balance must be struck regarding reconstruction fidelity and render speed, we have shown that anatomy representations can achieve useful quality levels on mobile and immersive hardware. 

\paragraph{Creation and Rendering}

Our experiments demonstrate that \ac{gs} can achieve high-quality anatomy visualization on mobile and immersive hardware, even for resource-constrained devices. The visual quality achieved is promising for anatomy rendering, corroborating the findings of Niedermayer et al. \cite{niedermayr_application_2024}. However, the computational demands are significant, particularly when aiming for optimal visual fidelity. Further quality improvements could be achieved through more sophisticated view selection algorithms. Optimizing for voxel visibility gain and view angle diversity can improve quality and generation time without significant runtime impact \cite{vazquez_representative_2008}. We opted for a more straightforward generation approach, rendering more viewpoints in this work as our focus was not on dataset generation. It would also be possible to significantly reduce the time to view new \ac{gs} visualizations from initial parameter selection by using more conservative parameters, fewer samples, denoising, or super-sampling. We tested our approach mainly on imagery generated from \ac{ct} data, but the approach should extend to other kinds of volumetric data. \Ac{ct} scans containing contrasting agents, multi-energy data from photon-counting \acp{ct} scanners, or magnetic resonance imaging data could be visualized with this approach, possibly with preparation steps like tissue segmentation.

One of the primary challenges in our work stems from the conflicting goals of maximizing visual quality, minimizing file size, and achieving interactive frame rates in rendering. Increasing the quality of the \ac{gs} anatomy representation is usually achieved by adjusting training parameters to increase the resulting Gaussian count. The increased Gaussian count leads to larger storage requirements and lower rendering performance. To mitigate this, we employ basic compression techniques and clustering. While our approach yields significant reductions in file size compared to industry-standard DICOM files and compressed volumes, further gains could be achieved by adopting more advanced methods from recent literature. Some compression techniques selectively retain relevant Gaussians, akin to our pruning approach. Combining these methods could further benefit mobile platforms by reducing Gaussian count, which is the leading performance bottleneck. Given the importance of Gaussian count in balancing quality and performance, finer control of their creation during the optimization process would be valuable. Currently, it's challenging to predict and adjust how many Gaussians will be used to encode a given volume.

\paragraph{Layering Approach}

Our layering approach has potential for bringing more interactivity to \ac{gs} and complex visualization to \ac{vr}. However, we identified some inefficiencies that the original 3DGS approach did not consider and that our approach initially exacerbated. To address this, we introduced an inactive pruning technique to remove superfluous Gaussians that don't significantly contribute to the final rendered image. While this approach reduces clutter and improves efficiency, it faces challenges in differentiating between inactive Gaussians due to being obscured by lower layers and those already optimally positioned. We used pruning with exponential falloff with the assumption that inactive Gaussians early in the optimization depend on information from the lower layers and can be removed. Still, alternative strategies might handle this trade-off differently. It should be noted that pruning also reduces the jarring experience of seeing unoptimized Gaussians while cutting through layers, which is not captured by our metrics.

The layering technique functions by encoding differences between layers, similar to how video codecs encode frame content as differences from previous frames. However, unlike video codecs, our current implementation doesn't allow upper layers to override, modify, or hide information from lower layers. Allowing upper layers to modify lower ones might offer more flexibility in representable information. Still, by avoiding interactions between layers, our current system ensures that each layer can be independently displayed and cut with manageable complexity.

A benefit of layering is that it generally results in smaller overall file sizes compared to training each anatomy view separately. However, this storage benefit comes with a potential increase in the number of Gaussians needed at runtime to display the same anatomy properly. The efficiency of the layered approach depends on how much of the lower layer is obscured by new information and how much can be reused. In anatomical data, we observed that adding muscles on top of bones doesn't result in significant storage savings, as most of the bone is obscured by muscle. However, adding a few blood vessels on top of muscle could be very lightweight. The global illumination effects inherent in path tracing complicate this, as even minor visual changes can affect the luminosity and color of nearby pixels, requiring additional Gaussians for minor visual changes. This challenge suggests that layered representations may be more situational in performance-constrained environments. Still, we observed solid performances on tested devices, suggesting that this approach is feasible for real-world scenarios.

Layering also introduces more interactivity to \ac{gs} rendering. The ability to cut to lower layers provides a reasonable visualization as long as the viewer doesn't look "inside" the cut layer, where optimization hasn't occurred. This opens opportunities for effects like fading between layers, which would be difficult to achieve with separate \ac{gs} representations.The visual quality of cutting edges depends on the Gaussian point cloud density at the edge. Smarter cutting algorithms, considering the full area covered by splats, could improve edge sharpness at a minor performance cost.

\subsection{Limitations}

While our layered \ac{gs} approach demonstrates significant potential for medical volumetric data visualization, it's important to acknowledge its current limitations. Although slicing layers is an improvement over static representations, there remains considerable room for enhancement to support full interactivity comparable to traditional volume rendering techniques. In cases where an upper layer obscures most of a lower one, both layers are required for accurate rendering, increasing the number of displayed splats. This might be addressed by enhancing the rendering or pruning process to remove redundant Gaussians from lower layers that do not contribute visually. Performance constraints also remain a consideration, particularly on standalone VR devices. While our approach is significantly more real-time capable than path tracing, the amount of detail we can include in a \ac{gs} representation that runs smoothly on these devices is limited. Zooming in on small structures, especially on low contrast regions, quickly reveals a lack of detail and splatting artifacts. We believe that the achievable detail is sufficient for many practical applications in medical visualization, but performance remains a bottleneck.

\subsection{Future Work}
Future research could focus on increasing the interactivity of \ac{gs}, especially while using a layered approach. More work is also needed in improving layer compression to avoid the performance cost of rendering all lower layers, or further refining the layering approach to ensure efficient encoding of new information. Additionally, the concept of layered \ac{gs} is not restricted to anatomy data. Inclusion of additional information that can be inferred from the existing data could also be explored, like depth maps for depth regularization. Future research could also explore its application in other domains with tight control over data, such as mechanical visualizations, industrial scans, or generative visualizations. These diverse applications could yield new insights and optimization techniques that could, in turn, benefit medical visualization.

\section{Conclusion}
We introduce a layered \ac{gs} reconstruction that can encode multiple visualizations of the same volume data, building upon each other similar to multiple transfer functions in traditional volume rendering. Extending the \ac{gs} adaptive control, we introduce inactive pruning to remove inactive Gaussians introduced in our layering approach. With this intermediate \ac{gs} representation, we can use high-quality medical volume rendering to otherwise compute-constrained scenarios, especially \ac{vr}. Our approach adds opportunities for interactivity to \ac{gs} rendering, such as the ability to cut through different layers, emulating how volume visualizations are currently used in medical practice. While still more experimentation and work are needed to bring full interactivity to \ac{gs}, we believe this approach can be useful for planning and educational situations where real-time path tracing is not viable or semi-static representations are sufficient. 

\acknowledgments{%
We want to thank Nikolai Hofmann for valuable insights regarding the used path tracer and d.hip for providing a campus stipend.
}

\section*{Supplementary Material}
Supplemental materials, generated training data, and trained models are available at \url{https://osf.io/tuwh5/}. Code is available at \url{https://github.com/roth-hex-lab/Multi-Layer-Anatomy-GS-Unity-Rendering}.
The original CT volumes used in this work are available from their respective authors.

\bibliographystyle{abbrv-doi-hyperref}

\bibliography{literature}

\begin{thebibliography}{10}

\bibitem{adinets_onesweep_2022}
\href{https://doi.org/10.48550/arXiv.2206.01784}{A.~Adinets and D.~Merrill}.
\newblock \href{https://doi.org/10.48550/arXiv.2206.01784}{Onesweep: {A}
  {Faster} {Least} {Significant} {Digit} {Radix} {Sort} for {GPUs}},
  \href{https://doi.org/10.48550/arXiv.2206.01784}{June 2022}.
  \href{https://doi.org/10.48550/arXiv.2206.01784}
{doi: {{%
10\hspace{.1pt}\discretionary{.}{%
}{.}\hspace{.4pt}48550\discretionary{/}{%
}{/}arXiv\hspace{.1pt}\discretionary{.}{%
}{.}\hspace{.4pt}2206\hspace{.1pt}\discretionary{.}{%
}{.}\hspace{.4pt}01784}}}


\bibitem{bagdasarian_3dgszip_2024}
\href{https://doi.org/10.48550/arXiv.2407.09510}{M.~T. Bagdasarian, P.~Knoll,
  F.~Barthel, A.~Hilsmann, P.~Eisert, and W.~Morgenstern}.
\newblock \href{https://doi.org/10.48550/arXiv.2407.09510}{{3DGS}.zip: {A}
  survey on {3D} {Gaussian} {Splatting} {Compression} {Methods}},
  \href{https://doi.org/10.48550/arXiv.2407.09510}{July 2024}.
  \href{https://doi.org/10.48550/arXiv.2407.09510}
{doi: {{%
10\hspace{.1pt}\discretionary{.}{%
}{.}\hspace{.4pt}48550\discretionary{/}{%
}{/}arXiv\hspace{.1pt}\discretionary{.}{%
}{.}\hspace{.4pt}2407\hspace{.1pt}\discretionary{.}{%
}{.}\hspace{.4pt}09510}}}


\bibitem{binder_leveraging_2019}
\href{https://doi.org/10.1016/j.aanat.2018.12.004}{J.~Binder, C.~Krautz,
  K.~Engel, R.~Grützmann, F.~A. Fellner, P.~H.~M. Burger, and M.~Scholz}.
\newblock \href{https://doi.org/10.1016/j.aanat.2018.12.004}{Leveraging medical
  imaging for medical education — {A} cinematic rendering-featured lecture}.
\newblock \href{https://doi.org/10.1016/j.aanat.2018.12.004}{{\em Annals of
  Anatomy - Anatomischer Anzeiger}},
  \href{https://doi.org/10.1016/j.aanat.2018.12.004}{222:159--165},
  \href{https://doi.org/10.1016/j.aanat.2018.12.004}{Mar. 2019}.
  \href{https://doi.org/10.1016/j.aanat.2018.12.004}
{doi: {{%
10\hspace{.1pt}\discretionary{.}{%
}{.}\hspace{.4pt}1016\discretionary{/}{%
}{/}j\hspace{.1pt}\discretionary{.}{%
}{.}\hspace{.4pt}aanat\hspace{.1pt}\discretionary{.}{%
}{.}\hspace{.4pt}2018\hspace{.1pt}\discretionary{.}{%
}{.}\hspace{.4pt}12\hspace{.1pt}\discretionary{.}{%
}{.}\hspace{.4pt}004}}}


\bibitem{binder_cinematic_2021}
\href{https://doi.org/10.1002/ase.1989}{J.~S. Binder, M.~Scholz, S.~Ellmann,
  M.~Uder, R.~Grützmann, G.~F. Weber, and C.~Krautz}.
\newblock \href{https://doi.org/10.1002/ase.1989}{Cinematic {Rendering} in
  {Anatomy}: {A} {Crossover} {Study} {Comparing} a {Novel} {3D}
  {Reconstruction} {Technique} to {Conventional} {Computed} {Tomography}}.
\newblock \href{https://doi.org/10.1002/ase.1989}{{\em Anatomical Sciences
  Education}}, \href{https://doi.org/10.1002/ase.1989}{14(1):22--31},
  \href{https://doi.org/10.1002/ase.1989}{2021}.
  \href{https://doi.org/10.1002/ase.1989}
{doi: {{%
10\hspace{.1pt}\discretionary{.}{%
}{.}\hspace{.4pt}1002\discretionary{/}{%
}{/}ase\hspace{.1pt}\discretionary{.}{%
}{.}\hspace{.4pt}1989}}}


\bibitem{cai_clnerf_2023}
\href{https://doi.org/10.1109/ICCV51070.2023.02119}{Z.~Cai and M.~Müller}.
\newblock \href{https://doi.org/10.1109/ICCV51070.2023.02119}{{CLNeRF}:
  {Continual} {Learning} {Meets} {NeRF}}.
\newblock \href{https://doi.org/10.1109/ICCV51070.2023.02119}{In {\em 2023
  {IEEE}/{CVF} {International} {Conference} on {Computer} {Vision} ({ICCV})}},
  \href{https://doi.org/10.1109/ICCV51070.2023.02119}{pp. 23128--23137}.
  \href{https://doi.org/10.1109/ICCV51070.2023.02119}{IEEE},
  \href{https://doi.org/10.1109/ICCV51070.2023.02119}{Paris, France},
  \href{https://doi.org/10.1109/ICCV51070.2023.02119}{Oct. 2023}.
  \href{https://doi.org/10.1109/ICCV51070.2023.02119}
{doi: {{%
10\hspace{.1pt}\discretionary{.}{%
}{.}\hspace{.4pt}1109\discretionary{/}{%
}{/}ICCV51070\hspace{.1pt}\discretionary{.}{%
}{.}\hspace{.4pt}2023\hspace{.1pt}\discretionary{.}{%
}{.}\hspace{.4pt}02119}}}


\bibitem{cardobi_path_2023}
\href{https://doi.org/10.3390/jimaging9020024}{N.~Cardobi, R.~Nocini,
  G.~Molteni, V.~Favero, A.~Fior, D.~Marchioni, S.~Montemezzi, and
  M.~D’Onofrio}.
\newblock \href{https://doi.org/10.3390/jimaging9020024}{Path {Tracing} vs.
  {Volume} {Rendering} {Technique} in {Post}-{Surgical} {Assessment} of {Bone}
  {Flap} in {Oncologic} {Head} and {Neck} {Reconstructive} {Surgery}: {A}
  {Preliminary} {Study}}.
\newblock \href{https://doi.org/10.3390/jimaging9020024}{{\em Journal of
  Imaging}}, \href{https://doi.org/10.3390/jimaging9020024}{9(2):24},
  \href{https://doi.org/10.3390/jimaging9020024}{Feb. 2023}.
  \href{https://doi.org/10.3390/jimaging9020024}
{doi: {{%
10\hspace{.1pt}\discretionary{.}{%
}{.}\hspace{.4pt}3390\discretionary{/}{%
}{/}jimaging9020024}}}


\bibitem{caton_jr_three-dimensional_2020}
\href{https://doi.org/10.1111/jon.12697}{M.~T. Caton~Jr., W.~F. Wiggins, and
  D.~Nunez}.
\newblock \href{https://doi.org/10.1111/jon.12697}{Three-{Dimensional}
  {Cinematic} {Rendering} to {Optimize} {Visualization} of {Cerebrovascular}
  {Anatomy} and {Disease} in {CT} {Angiography}}.
\newblock \href{https://doi.org/10.1111/jon.12697}{{\em Journal of
  Neuroimaging}}, \href{https://doi.org/10.1111/jon.12697}{30(3):286--296},
  \href{https://doi.org/10.1111/jon.12697}{2020}.
  \href{https://doi.org/10.1111/jon.12697}
{doi: {{%
10\hspace{.1pt}\discretionary{.}{%
}{.}\hspace{.4pt}1111\discretionary{/}{%
}{/}jon\hspace{.1pt}\discretionary{.}{%
}{.}\hspace{.4pt}12697}}}


\bibitem{chen_hac_2024}
\href{https://doi.org/10.48550/arXiv.2403.14530}{Y.~Chen, Q.~Wu, W.~Lin,
  M.~Harandi, and J.~Cai}.
\newblock \href{https://doi.org/10.48550/arXiv.2403.14530}{{HAC}: {Hash}-grid
  {Assisted} {Context} for {3D} {Gaussian} {Splatting} {Compression}},
  \href{https://doi.org/10.48550/arXiv.2403.14530}{July 2024}.
  \href{https://doi.org/10.48550/arXiv.2403.14530}
{doi: {{%
10\hspace{.1pt}\discretionary{.}{%
}{.}\hspace{.4pt}48550\discretionary{/}{%
}{/}arXiv\hspace{.1pt}\discretionary{.}{%
}{.}\hspace{.4pt}2403\hspace{.1pt}\discretionary{.}{%
}{.}\hspace{.4pt}14530}}}


\bibitem{chung_meil-nerf_2022}
\href{https://doi.org/10.48550/arXiv.2212.08328}{J.~Chung, K.~Lee, S.~Baik, and
  K.~M. Lee}.
\newblock \href{https://doi.org/10.48550/arXiv.2212.08328}{{MEIL}-{NeRF}:
  {Memory}-{Efficient} {Incremental} {Learning} of {Neural} {Radiance}
  {Fields}}, \href{https://doi.org/10.48550/arXiv.2212.08328}{Dec. 2022}.
  \href{https://doi.org/10.48550/arXiv.2212.08328}
{doi: {{%
10\hspace{.1pt}\discretionary{.}{%
}{.}\hspace{.4pt}48550\discretionary{/}{%
}{/}arXiv\hspace{.1pt}\discretionary{.}{%
}{.}\hspace{.4pt}2212\hspace{.1pt}\discretionary{.}{%
}{.}\hspace{.4pt}08328}}}


\bibitem{comaniciu_shaping_2016}
\href{https://doi.org/10.1016/j.media.2016.06.016}{D.~Comaniciu, K.~Engel,
  B.~Georgescu, and T.~Mansi}.
\newblock \href{https://doi.org/10.1016/j.media.2016.06.016}{Shaping the future
  through innovations: {From} medical imaging to precision medicine}.
\newblock \href{https://doi.org/10.1016/j.media.2016.06.016}{{\em Medical Image
  Analysis}}, \href{https://doi.org/10.1016/j.media.2016.06.016}{33:19--26},
  \href{https://doi.org/10.1016/j.media.2016.06.016}{Oct. 2016}.
  \href{https://doi.org/10.1016/j.media.2016.06.016}
{doi: {{%
10\hspace{.1pt}\discretionary{.}{%
}{.}\hspace{.4pt}1016\discretionary{/}{%
}{/}j\hspace{.1pt}\discretionary{.}{%
}{.}\hspace{.4pt}media\hspace{.1pt}\discretionary{.}{%
}{.}\hspace{.4pt}2016\hspace{.1pt}\discretionary{.}{%
}{.}\hspace{.4pt}06\hspace{.1pt}\discretionary{.}{%
}{.}\hspace{.4pt}016}}}


\bibitem{dappa_cinematic_2016}
\href{https://doi.org/10.1007/s13244-016-0518-1}{E.~Dappa, K.~Higashigaito,
  J.~Fornaro, S.~Leschka, S.~Wildermuth, and H.~Alkadhi}.
\newblock \href{https://doi.org/10.1007/s13244-016-0518-1}{Cinematic rendering
  – an alternative to volume rendering for {3D} computed tomography imaging}.
\newblock \href{https://doi.org/10.1007/s13244-016-0518-1}{{\em Insights into
  Imaging}}, \href{https://doi.org/10.1007/s13244-016-0518-1}{7(6):849--856},
  \href{https://doi.org/10.1007/s13244-016-0518-1}{Dec. 2016}.
  \href{https://doi.org/10.1007/s13244-016-0518-1}
{doi: {{%
10\hspace{.1pt}\discretionary{.}{%
}{.}\hspace{.4pt}1007\discretionary{/}{%
}{/}s13244\discretionary{%
}{-}{-}016\discretionary{%
}{-}{-}0518\discretionary{%
}{-}{-}1}}}


\bibitem{deng_fov-nerf_2022}
\href{https://doi.org/10.1109/TVCG.2022.3203102}{N.~Deng, Z.~He, J.~Ye,
  B.~Duinkharjav, P.~Chakravarthula, X.~Yang, and Q.~Sun}.
\newblock \href{https://doi.org/10.1109/TVCG.2022.3203102}{{FoV}-{NeRF}:
  {Foveated} {Neural} {Radiance} {Fields} for {Virtual} {Reality}}.
\newblock \href{https://doi.org/10.1109/TVCG.2022.3203102}{{\em IEEE
  Transactions on Visualization and Computer Graphics}},
  \href{https://doi.org/10.1109/TVCG.2022.3203102}{28(11):3854--3864},
  \href{https://doi.org/10.1109/TVCG.2022.3203102}{Nov. 2022}.
  \href{https://doi.org/10.1109/TVCG.2022.3203102}
{doi: {{%
10\hspace{.1pt}\discretionary{.}{%
}{.}\hspace{.4pt}1109\discretionary{/}{%
}{/}TVCG\hspace{.1pt}\discretionary{.}{%
}{.}\hspace{.4pt}2022\hspace{.1pt}\discretionary{.}{%
}{.}\hspace{.4pt}3203102}}}


\bibitem{engel_real-time_2006}
\href{https://doi.org/10.2312/egt.20061064}{K.~Engel, M.~Hadwiger, J.~M. Kniss,
  and C.~Rezk-Salama}.
\newblock \href{https://doi.org/10.2312/egt.20061064}{{Real-Time Volume
  Graphics}}.
\newblock \href{https://doi.org/10.2312/egt.20061064}{In N.~Magnenat-Thalmann
  and K.~Bühler, eds., {\em Eurographics 2006: Tutorials}}.
  \href{https://doi.org/10.2312/egt.20061064}{The Eurographics Association},
  \href{https://doi.org/10.2312/egt.20061064}{2006}.
  \href{https://doi.org/10.2312/egt.20061064}
{doi: {{%
10\hspace{.1pt}\discretionary{.}{%
}{.}\hspace{.4pt}2312\discretionary{/}{%
}{/}egt\hspace{.1pt}\discretionary{.}{%
}{.}\hspace{.4pt}20061064}}}


\bibitem{frieder_back--front_1985}
\href{https://doi.org/10.1109/MCG.1985.276273}{G.~Frieder, D.~Gordon, and
  R.~Reynolds}.
\newblock \href{https://doi.org/10.1109/MCG.1985.276273}{Back-to-{Front}
  {Display} of {Voxel} {Based} {Objects}}.
\newblock \href{https://doi.org/10.1109/MCG.1985.276273}{{\em IEEE Computer
  Graphics and Applications}},
  \href{https://doi.org/10.1109/MCG.1985.276273}{5(1):52--60},
  \href{https://doi.org/10.1109/MCG.1985.276273}{1985}.
  \href{https://doi.org/10.1109/MCG.1985.276273}
{doi: {{%
10\hspace{.1pt}\discretionary{.}{%
}{.}\hspace{.4pt}1109\discretionary{/}{%
}{/}MCG\hspace{.1pt}\discretionary{.}{%
}{.}\hspace{.4pt}1985\hspace{.1pt}\discretionary{.}{%
}{.}\hspace{.4pt}276273}}}


\bibitem{glemser_new_2018}
\href{https://doi.org/10.1016/j.wneu.2018.02.174}{P.~A. Glemser, K.~Engel,
  D.~Simons, J.~Steffens, H.-P. Schlemmer, and B.~Orakcioglu}.
\newblock \href{https://doi.org/10.1016/j.wneu.2018.02.174}{A {New} {Approach}
  for {Photorealistic} {Visualization} of {Rendered} {Computed}
  {Tomography} {Images}}.
\newblock \href{https://doi.org/10.1016/j.wneu.2018.02.174}{{\em World
  Neurosurgery}},
  \href{https://doi.org/10.1016/j.wneu.2018.02.174}{114:e283--e292},
  \href{https://doi.org/10.1016/j.wneu.2018.02.174}{June 2018}.
  \href{https://doi.org/10.1016/j.wneu.2018.02.174}
{doi: {{%
10\hspace{.1pt}\discretionary{.}{%
}{.}\hspace{.4pt}1016\discretionary{/}{%
}{/}j\hspace{.1pt}\discretionary{.}{%
}{.}\hspace{.4pt}wneu\hspace{.1pt}\discretionary{.}{%
}{.}\hspace{.4pt}2018\hspace{.1pt}\discretionary{.}{%
}{.}\hspace{.4pt}02\hspace{.1pt}\discretionary{.}{%
}{.}\hspace{.4pt}174}}}


\bibitem{hofmann_efficient_2021}
\href{https://doi.org/10.1007/978-1-4842-7185-8_43}{N.~Hofmann and A.~Evans}.
\newblock \href{https://doi.org/10.1007/978-1-4842-7185-8_43}{Efficient
  {Unbiased} {Volume} {Path} {Tracing} on the {GPU}}.
\newblock \href{https://doi.org/10.1007/978-1-4842-7185-8_43}{In A.~Marrs,
  P.~Shirley, and I.~Wald, eds., {\em Ray {Tracing} {Gems} {II}: {Next}
  {Generation} {Real}-{Time} {Rendering} with {DXR}, {Vulkan}, and {OptiX}}},
  \href{https://doi.org/10.1007/978-1-4842-7185-8_43}{pp. 699--711}.
  \href{https://doi.org/10.1007/978-1-4842-7185-8_43}{Apress},
  \href{https://doi.org/10.1007/978-1-4842-7185-8_43}{Berkeley, CA},
  \href{https://doi.org/10.1007/978-1-4842-7185-8_43}{2021}.
  \href{https://doi.org/10.1007/978-1-4842-7185-8_43}
{doi: {{%
10\hspace{.1pt}\discretionary{.}{%
}{.}\hspace{.4pt}1007\discretionary{/}{%
}{/}978\discretionary{%
}{-}{-}1\discretionary{%
}{-}{-}4842\discretionary{%
}{-}{-}7185\discretionary{%
}{-}{-}8\_43}}}


\bibitem{hofmann_neural_2020}
\href{https://doi.org/10.1145/3406181}{N.~Hofmann, J.~Martschinke, K.~Engel,
  and M.~Stamminger}.
\newblock \href{https://doi.org/10.1145/3406181}{Neural {Denoising} for {Path}
  {Tracing} of {Medical} {Volumetric} {Data}}.
\newblock \href{https://doi.org/10.1145/3406181}{{\em Proceedings of the ACM on
  Computer Graphics and Interactive Techniques}},
  \href{https://doi.org/10.1145/3406181}{3(2):1--18},
  \href{https://doi.org/10.1145/3406181}{Aug. 2020}.
  \href{https://doi.org/10.1145/3406181}
{doi: {{%
10\hspace{.1pt}\discretionary{.}{%
}{.}\hspace{.4pt}1145\discretionary{/}{%
}{/}3406181}}}


\bibitem{huo_survey_2021}
\href{https://doi.org/10.1007/s41095-021-0209-9}{Y.~Huo and S.-e. Yoon}.
\newblock \href{https://doi.org/10.1007/s41095-021-0209-9}{A survey on deep
  learning-based {Monte} {Carlo} denoising}.
\newblock \href{https://doi.org/10.1007/s41095-021-0209-9}{{\em Computational
  Visual Media}},
  \href{https://doi.org/10.1007/s41095-021-0209-9}{7(2):169--185},
  \href{https://doi.org/10.1007/s41095-021-0209-9}{June 2021}.
  \href{https://doi.org/10.1007/s41095-021-0209-9}
{doi: {{%
10\hspace{.1pt}\discretionary{.}{%
}{.}\hspace{.4pt}1007\discretionary{/}{%
}{/}s41095\discretionary{%
}{-}{-}021\discretionary{%
}{-}{-}0209\discretionary{%
}{-}{-}9}}}


\bibitem{iglesias-guitian_real-time_2022}
\href{https://doi.org/10.1109/TVCG.2020.3037680}{J.~A. Iglesias-Guitian,
  P.~Mane, and B.~Moon}.
\newblock \href{https://doi.org/10.1109/TVCG.2020.3037680}{Real-{Time}
  {Denoising} of {Volumetric} {Path} {Tracing} for {Direct} {Volume}
  {Rendering}}.
\newblock \href{https://doi.org/10.1109/TVCG.2020.3037680}{{\em IEEE
  Transactions on Visualization and Computer Graphics}},
  \href{https://doi.org/10.1109/TVCG.2020.3037680}{28(7):2734--2747},
  \href{https://doi.org/10.1109/TVCG.2020.3037680}{July 2022}.
  \href{https://doi.org/10.1109/TVCG.2020.3037680}
{doi: {{%
10\hspace{.1pt}\discretionary{.}{%
}{.}\hspace{.4pt}1109\discretionary{/}{%
}{/}TVCG\hspace{.1pt}\discretionary{.}{%
}{.}\hspace{.4pt}2020\hspace{.1pt}\discretionary{.}{%
}{.}\hspace{.4pt}3037680}}}


\bibitem{jiang_vr-gs_2024}
\href{https://doi.org/10.1145/3641519.3657448}{Y.~Jiang, C.~Yu, T.~Xie, X.~Li,
  Y.~Feng, H.~Wang, M.~Li, H.~Lau, F.~Gao, Y.~Yang, and C.~Jiang}.
\newblock \href{https://doi.org/10.1145/3641519.3657448}{{VR}-{GS}: {A}
  {Physical} {Dynamics}-{Aware} {Interactive} {Gaussian} {Splatting} {System}
  in {Virtual} {Reality}}.
\newblock \href{https://doi.org/10.1145/3641519.3657448}{In {\em {ACM}
  {SIGGRAPH} 2024 {Conference} {Papers}}},
  \href{https://doi.org/10.1145/3641519.3657448}{{SIGGRAPH} '24},
  \href{https://doi.org/10.1145/3641519.3657448}{p.~1}.
  \href{https://doi.org/10.1145/3641519.3657448}{Association for Computing
  Machinery}, \href{https://doi.org/10.1145/3641519.3657448}{New York, NY,
  USA}, \href{https://doi.org/10.1145/3641519.3657448}{July 2024}.
  \href{https://doi.org/10.1145/3641519.3657448}
{doi: {{%
10\hspace{.1pt}\discretionary{.}{%
}{.}\hspace{.4pt}1145\discretionary{/}{%
}{/}3641519\hspace{.1pt}\discretionary{.}{%
}{.}\hspace{.4pt}3657448}}}


\bibitem{kerbl3Dgaussians}
\href{https://repo-sam.inria.fr/fungraph/3d-gaussian-splatting/}{B.~Kerbl,
  G.~Kopanas, T.~Leimk{\"u}hler, and G.~Drettakis}.
\newblock \href{https://repo-sam.inria.fr/fungraph/3d-gaussian-splatting/}{3d
  gaussian splatting for real-time radiance field rendering}.
\newblock \href{https://repo-sam.inria.fr/fungraph/3d-gaussian-splatting/}{{\em
  ACM Transactions on Graphics}},
  \href{https://repo-sam.inria.fr/fungraph/3d-gaussian-splatting/}{42(4)},
  \href{https://repo-sam.inria.fr/fungraph/3d-gaussian-splatting/}{July 2023}.

\bibitem{kerbl_hierarchical_2024}
\href{https://doi.org/10.1145/3658160}{B.~Kerbl, A.~Meuleman, G.~Kopanas,
  M.~Wimmer, A.~Lanvin, and G.~Drettakis}.
\newblock \href{https://doi.org/10.1145/3658160}{A {Hierarchical} {3D}
  {Gaussian} {Representation} for {Real}-{Time} {Rendering} of {Very} {Large}
  {Datasets}}.
\newblock \href{https://doi.org/10.1145/3658160}{{\em ACM Transactions on
  Graphics}}, \href{https://doi.org/10.1145/3658160}{43(4):1--15},
  \href{https://doi.org/10.1145/3658160}{July 2024}.
  \href{https://doi.org/10.1145/3658160}
{doi: {{%
10\hspace{.1pt}\discretionary{.}{%
}{.}\hspace{.4pt}1145\discretionary{/}{%
}{/}3658160}}}


\bibitem{kleinbeck_adaptive_2023}
\href{https://doi.org/10.1109/ISMAR-Adjunct60411.2023.00131}{C.~Kleinbeck,
  M.~Smietana, N.~Lewis, T.~Teufel, J.~Kreimeier, C.~Heinzl, J.~Steiner,
  C.~Anthes, and D.~Roth}.
\newblock \href{https://doi.org/10.1109/ISMAR-Adjunct60411.2023.00131}{Adaptive
  {Volumetric} {Anatomy} {Visualization} in {VR} with {Tangible} {Control}}.
\newblock \href{https://doi.org/10.1109/ISMAR-Adjunct60411.2023.00131}{In {\em
  2023 {IEEE} {International} {Symposium} on {Mixed} and {Augmented} {Reality}
  {Adjunct} ({ISMAR}-{Adjunct})}},
  \href{https://doi.org/10.1109/ISMAR-Adjunct60411.2023.00131}{pp. 613--614},
  \href{https://doi.org/10.1109/ISMAR-Adjunct60411.2023.00131}{Oct. 2023}.
  \href{https://doi.org/10.1109/ISMAR-Adjunct60411.2023.00131}
{doi: {{%
10\hspace{.1pt}\discretionary{.}{%
}{.}\hspace{.4pt}1109\discretionary{/}{%
}{/}ISMAR\discretionary{%
}{-}{-}Adjunct60411\hspace{.1pt}\discretionary{.}{%
}{.}\hspace{.4pt}2023\hspace{.1pt}\discretionary{.}{%
}{.}\hspace{.4pt}00131}}}


\bibitem{kleinbeck_neural_2024}
\href{https://doi.org/10.1007/s11548-024-03143-w}{C.~Kleinbeck, H.~Zhang, B.~D.
  Killeen, D.~Roth, and M.~Unberath}.
\newblock \href{https://doi.org/10.1007/s11548-024-03143-w}{Neural digital
  twins: reconstructing complex medical environments for spatial planning in
  virtual reality}.
\newblock \href{https://doi.org/10.1007/s11548-024-03143-w}{{\em International
  Journal of Computer Assisted Radiology and Surgery}},
  \href{https://doi.org/10.1007/s11548-024-03143-w}{May 2024}.
  \href{https://doi.org/10.1007/s11548-024-03143-w}
{doi: {{%
10\hspace{.1pt}\discretionary{.}{%
}{.}\hspace{.4pt}1007\discretionary{/}{%
}{/}s11548\discretionary{%
}{-}{-}024\discretionary{%
}{-}{-}03143\discretionary{%
}{-}{-}w}}}


\bibitem{kroes_exposure_2012}
\href{https://doi.org/10.1371/journal.pone.0038586}{T.~Kroes, F.~H. Post, and
  C.~P. Botha}.
\newblock \href{https://doi.org/10.1371/journal.pone.0038586}{Exposure
  {Render}: {An} {Interactive} {Photo}-{Realistic} {Volume} {Rendering}
  {Framework}}.
\newblock \href{https://doi.org/10.1371/journal.pone.0038586}{{\em PLOS ONE}},
  \href{https://doi.org/10.1371/journal.pone.0038586}{7(7):e38586},
  \href{https://doi.org/10.1371/journal.pone.0038586}{July 2012}.
  \href{https://doi.org/10.1371/journal.pone.0038586}
{doi: {{%
10\hspace{.1pt}\discretionary{.}{%
}{.}\hspace{.4pt}1371\discretionary{/}{%
}{/}journal\hspace{.1pt}\discretionary{.}{%
}{.}\hspace{.4pt}pone\hspace{.1pt}\discretionary{.}{%
}{.}\hspace{.4pt}0038586}}}


\bibitem{levoy_display_1988}
\href{https://doi.org/10.1109/38.511}{M.~Levoy}.
\newblock \href{https://doi.org/10.1109/38.511}{Display of surfaces from volume
  data}.
\newblock \href{https://doi.org/10.1109/38.511}{{\em IEEE Computer Graphics and
  Applications}}, \href{https://doi.org/10.1109/38.511}{8(3):29--37},
  \href{https://doi.org/10.1109/38.511}{May 1988}.
  \href{https://doi.org/10.1109/38.511}
{doi: {{%
10\hspace{.1pt}\discretionary{.}{%
}{.}\hspace{.4pt}1109\discretionary{/}{%
}{/}38\hspace{.1pt}\discretionary{.}{%
}{.}\hspace{.4pt}511}}}


\bibitem{levoy_light_1996}
\href{https://doi.org/10.1145/237170.237199}{M.~Levoy and P.~Hanrahan}.
\newblock \href{https://doi.org/10.1145/237170.237199}{Light field rendering}.
\newblock \href{https://doi.org/10.1145/237170.237199}{In {\em Proceedings of
  the 23rd annual conference on {Computer} graphics and interactive
  techniques}}, \href{https://doi.org/10.1145/237170.237199}{{SIGGRAPH} '96},
  \href{https://doi.org/10.1145/237170.237199}{pp. 31--42}.
  \href{https://doi.org/10.1145/237170.237199}{Association for Computing
  Machinery}, \href{https://doi.org/10.1145/237170.237199}{New York, NY, USA},
  \href{https://doi.org/10.1145/237170.237199}{Aug. 1996}.
  \href{https://doi.org/10.1145/237170.237199}
{doi: {{%
10\hspace{.1pt}\discretionary{.}{%
}{.}\hspace{.4pt}1145\discretionary{/}{%
}{/}237170\hspace{.1pt}\discretionary{.}{%
}{.}\hspace{.4pt}237199}}}


\bibitem{li_rt-nerf_2022}
\href{https://doi.org/10.1145/3508352.3549380}{C.~Li, S.~Li, Y.~Zhao, W.~Zhu,
  and Y.~Lin}.
\newblock \href{https://doi.org/10.1145/3508352.3549380}{{RT}-{NeRF}:
  {Real}-{Time} {On}-{Device} {Neural} {Radiance} {Fields} {Towards}
  {Immersive} {AR}/{VR} {Rendering}}.
\newblock \href{https://doi.org/10.1145/3508352.3549380}{In {\em Proceedings of
  the 41st {IEEE}/{ACM} {International} {Conference} on {Computer}-{Aided}
  {Design}}}, \href{https://doi.org/10.1145/3508352.3549380}{{ICCAD} '22},
  \href{https://doi.org/10.1145/3508352.3549380}{pp. 1--9}.
  \href{https://doi.org/10.1145/3508352.3549380}{Association for Computing
  Machinery}, \href{https://doi.org/10.1145/3508352.3549380}{New York, NY,
  USA}, \href{https://doi.org/10.1145/3508352.3549380}{Dec. 2022}.
  \href{https://doi.org/10.1145/3508352.3549380}
{doi: {{%
10\hspace{.1pt}\discretionary{.}{%
}{.}\hspace{.4pt}1145\discretionary{/}{%
}{/}3508352\hspace{.1pt}\discretionary{.}{%
}{.}\hspace{.4pt}3549380}}}


\bibitem{li_compressing_2022}
\href{https://doi.org/10.48550/arXiv.2211.16386}{L.~Li, Z.~Shen, Z.~Wang,
  L.~Shen, and L.~Bo}.
\newblock \href{https://doi.org/10.48550/arXiv.2211.16386}{Compressing
  {Volumetric} {Radiance} {Fields} to 1 {MB}},
  \href{https://doi.org/10.48550/arXiv.2211.16386}{Nov. 2022}.
  \href{https://doi.org/10.48550/arXiv.2211.16386}
{doi: {{%
10\hspace{.1pt}\discretionary{.}{%
}{.}\hspace{.4pt}48550\discretionary{/}{%
}{/}arXiv\hspace{.1pt}\discretionary{.}{%
}{.}\hspace{.4pt}2211\hspace{.1pt}\discretionary{.}{%
}{.}\hspace{.4pt}16386}}}


\bibitem{liu2024lgslightweight4dGaussian}
\href{https://arxiv.org/abs/2406.16073}{H.~Liu, Y.~Liu, C.~Li, W.~Li, and
  Y.~Yuan}.
\newblock \href{https://arxiv.org/abs/2406.16073}{Lgs: A light-weight 4d
  gaussian splatting for efficient surgical scene reconstruction},
  \href{https://arxiv.org/abs/2406.16073}{2024}.

\bibitem{liu_endogaussian_2024}
\href{https://doi.org/10.48550/arXiv.2401.12561}{Y.~Liu, C.~Li, C.~Yang, and
  Y.~Yuan}.
\newblock \href{https://doi.org/10.48550/arXiv.2401.12561}{{EndoGaussian}:
  {Real}-time {Gaussian} {Splatting} for {Dynamic} {Endoscopic} {Scene}
  {Reconstruction}}, \href{https://doi.org/10.48550/arXiv.2401.12561}{Feb.
  2024}. \href{https://doi.org/10.48550/arXiv.2401.12561}
{doi: {{%
10\hspace{.1pt}\discretionary{.}{%
}{.}\hspace{.4pt}48550\discretionary{/}{%
}{/}arXiv\hspace{.1pt}\discretionary{.}{%
}{.}\hspace{.4pt}2401\hspace{.1pt}\discretionary{.}{%
}{.}\hspace{.4pt}12561}}}


\bibitem{ljung_state_2016}
\href{https://doi.org/10.1111/cgf.12934}{P.~Ljung, J.~Krüger, E.~Groller,
  M.~Hadwiger, C.~D. Hansen, and A.~Ynnerman}.
\newblock \href{https://doi.org/10.1111/cgf.12934}{State of the {Art} in
  {Transfer} {Functions} for {Direct} {Volume} {Rendering}}.
\newblock \href{https://doi.org/10.1111/cgf.12934}{{\em Computer Graphics
  Forum}}, \href{https://doi.org/10.1111/cgf.12934}{35(3):669--691},
  \href{https://doi.org/10.1111/cgf.12934}{2016}.
  \href{https://doi.org/10.1111/cgf.12934}
{doi: {{%
10\hspace{.1pt}\discretionary{.}{%
}{.}\hspace{.4pt}1111\discretionary{/}{%
}{/}cgf\hspace{.1pt}\discretionary{.}{%
}{.}\hspace{.4pt}12934}}}


\bibitem{mallya_packnet_2018}
\href{https://doi.org/10.1109/CVPR.2018.00810}{A.~Mallya and S.~Lazebnik}.
\newblock \href{https://doi.org/10.1109/CVPR.2018.00810}{{PackNet}: {Adding}
  {Multiple} {Tasks} to a {Single} {Network} by {Iterative} {Pruning}}.
\newblock \href{https://doi.org/10.1109/CVPR.2018.00810}{In {\em 2018
  {IEEE}/{CVF} {Conference} on {Computer} {Vision} and {Pattern}
  {Recognition}}}, \href{https://doi.org/10.1109/CVPR.2018.00810}{pp.
  7765--7773}, \href{https://doi.org/10.1109/CVPR.2018.00810}{June 2018}.
  \href{https://doi.org/10.1109/CVPR.2018.00810}
{doi: {{%
10\hspace{.1pt}\discretionary{.}{%
}{.}\hspace{.4pt}1109\discretionary{/}{%
}{/}CVPR\hspace{.1pt}\discretionary{.}{%
}{.}\hspace{.4pt}2018\hspace{.1pt}\discretionary{.}{%
}{.}\hspace{.4pt}00810}}}


\bibitem{mccloskey_catastrophic_1989}
\href{https://doi.org/10.1016/S0079-7421(08)60536-8}{M.~McCloskey and N.~J.
  Cohen}.
\newblock \href{https://doi.org/10.1016/S0079-7421(08)60536-8}{Catastrophic
  {Interference} in {Connectionist} {Networks}: {The} {Sequential} {Learning}
  {Problem}}.
\newblock \href{https://doi.org/10.1016/S0079-7421(08)60536-8}{In G.~H. Bower,
  ed., {\em Psychology of {Learning} and {Motivation}}},
  \href{https://doi.org/10.1016/S0079-7421(08)60536-8}{vol.~24},
  \href{https://doi.org/10.1016/S0079-7421(08)60536-8}{pp. 109--165}.
  \href{https://doi.org/10.1016/S0079-7421(08)60536-8}{Academic Press},
  \href{https://doi.org/10.1016/S0079-7421(08)60536-8}{Jan. 1989}.
  \href{https://doi.org/10.1016/S0079-7421(08)60536-8}
{doi: {{%
10\hspace{.1pt}\discretionary{.}{%
}{.}\hspace{.4pt}1016\discretionary{/}{%
}{/}S0079\discretionary{%
}{-}{-}7421\discretionary{%
}{(}{(}08\discretionary{)}{%
}{)}60536\discretionary{%
}{-}{-}8}}}


\bibitem{mildenhall_nerf_2020}
\href{https://doi.org/10.48550/arXiv.2003.08934}{B.~Mildenhall, P.~P.
  Srinivasan, M.~Tancik, J.~T. Barron, R.~Ramamoorthi, and R.~Ng}.
\newblock \href{https://doi.org/10.48550/arXiv.2003.08934}{{NeRF}:
  {Representing} {Scenes} as {Neural} {Radiance} {Fields} for {View}
  {Synthesis}}, \href{https://doi.org/10.48550/arXiv.2003.08934}{Aug. 2020}.
  \href{https://doi.org/10.48550/arXiv.2003.08934}
{doi: {{%
10\hspace{.1pt}\discretionary{.}{%
}{.}\hspace{.4pt}48550\discretionary{/}{%
}{/}arXiv\hspace{.1pt}\discretionary{.}{%
}{.}\hspace{.4pt}2003\hspace{.1pt}\discretionary{.}{%
}{.}\hspace{.4pt}08934}}}


\bibitem{morgenstern_compact_2024}
\href{https://doi.org/10.48550/arXiv.2312.13299}{W.~Morgenstern, F.~Barthel,
  A.~Hilsmann, and P.~Eisert}.
\newblock \href{https://doi.org/10.48550/arXiv.2312.13299}{Compact {3D} {Scene}
  {Representation} via {Self}-{Organizing} {Gaussian} {Grids}},
  \href{https://doi.org/10.48550/arXiv.2312.13299}{May 2024}.
  \href{https://doi.org/10.48550/arXiv.2312.13299}
{doi: {{%
10\hspace{.1pt}\discretionary{.}{%
}{.}\hspace{.4pt}48550\discretionary{/}{%
}{/}arXiv\hspace{.1pt}\discretionary{.}{%
}{.}\hspace{.4pt}2312\hspace{.1pt}\discretionary{.}{%
}{.}\hspace{.4pt}13299}}}


\bibitem{muller_instant_2022}
\href{https://doi.org/10.1145/3528223.3530127}{T.~Müller, A.~Evans, C.~Schied,
  and A.~Keller}.
\newblock \href{https://doi.org/10.1145/3528223.3530127}{Instant neural
  graphics primitives with a multiresolution hash encoding}.
\newblock \href{https://doi.org/10.1145/3528223.3530127}{{\em ACM Trans.
  Graph.}},
  \href{https://doi.org/10.1145/3528223.3530127}{41(4):102:1--102:15},
  \href{https://doi.org/10.1145/3528223.3530127}{July 2022}.
  \href{https://doi.org/10.1145/3528223.3530127}
{doi: {{%
10\hspace{.1pt}\discretionary{.}{%
}{.}\hspace{.4pt}1145\discretionary{/}{%
}{/}3528223\hspace{.1pt}\discretionary{.}{%
}{.}\hspace{.4pt}3530127}}}


\bibitem{navaneet_compact3d_2024}
\href{https://doi.org/10.48550/arXiv.2311.18159}{K.~L. Navaneet, K.~P. Meibodi,
  S.~A. Koohpayegani, and H.~Pirsiavash}.
\newblock \href{https://doi.org/10.48550/arXiv.2311.18159}{{Compact3D}:
  {Smaller} and {Faster} {Gaussian} {Splatting} with {Vector} {Quantization}},
  \href{https://doi.org/10.48550/arXiv.2311.18159}{June 2024}.
  \href{https://doi.org/10.48550/arXiv.2311.18159}
{doi: {{%
10\hspace{.1pt}\discretionary{.}{%
}{.}\hspace{.4pt}48550\discretionary{/}{%
}{/}arXiv\hspace{.1pt}\discretionary{.}{%
}{.}\hspace{.4pt}2311\hspace{.1pt}\discretionary{.}{%
}{.}\hspace{.4pt}18159}}}


\bibitem{niedermayr_application_2024}
\href{https://doi.org/10.48550/arXiv.2404.11285}{S.~Niedermayr, C.~Neuhauser,
  K.~Petkov, K.~Engel, and R.~Westermann}.
\newblock \href{https://doi.org/10.48550/arXiv.2404.11285}{Application of {3D}
  {Gaussian} {Splatting} for {Cinematic} {Anatomy} on {Consumer} {Class}
  {Devices}}, \href{https://doi.org/10.48550/arXiv.2404.11285}{June 2024}.
  \href{https://doi.org/10.48550/arXiv.2404.11285}
{doi: {{%
10\hspace{.1pt}\discretionary{.}{%
}{.}\hspace{.4pt}48550\discretionary{/}{%
}{/}arXiv\hspace{.1pt}\discretionary{.}{%
}{.}\hspace{.4pt}2404\hspace{.1pt}\discretionary{.}{%
}{.}\hspace{.4pt}11285}}}


\bibitem{po_instant_2023}
\href{https://doi.org/10.1109/ICCVW60793.2023.00358}{R.~Po, Z.~Dong, A.~W.
  Bergman, and G.~Wetzstein}.
\newblock \href{https://doi.org/10.1109/ICCVW60793.2023.00358}{Instant
  {Continual} {Learning} of {Neural} {Radiance} {Fields}}.
\newblock \href{https://doi.org/10.1109/ICCVW60793.2023.00358}{In {\em 2023
  {IEEE}/{CVF} {International} {Conference} on {Computer} {Vision} {Workshops}
  ({ICCVW})}}, \href{https://doi.org/10.1109/ICCVW60793.2023.00358}{pp.
  3326--3336}, \href{https://doi.org/10.1109/ICCVW60793.2023.00358}{Oct. 2023}.
  \href{https://doi.org/10.1109/ICCVW60793.2023.00358}
{doi: {{%
10\hspace{.1pt}\discretionary{.}{%
}{.}\hspace{.4pt}1109\discretionary{/}{%
}{/}ICCVW60793\hspace{.1pt}\discretionary{.}{%
}{.}\hspace{.4pt}2023\hspace{.1pt}\discretionary{.}{%
}{.}\hspace{.4pt}00358}}}


\bibitem{preim_survey_2018}
\href{https://doi.org/10.1016/j.cag.2018.01.005}{B.~Preim and P.~Saalfeld}.
\newblock \href{https://doi.org/10.1016/j.cag.2018.01.005}{A survey of virtual
  human anatomy education systems}.
\newblock \href{https://doi.org/10.1016/j.cag.2018.01.005}{{\em Computers \&
  Graphics}}, \href{https://doi.org/10.1016/j.cag.2018.01.005}{71:132--153},
  \href{https://doi.org/10.1016/j.cag.2018.01.005}{Apr. 2018}.
  \href{https://doi.org/10.1016/j.cag.2018.01.005}
{doi: {{%
10\hspace{.1pt}\discretionary{.}{%
}{.}\hspace{.4pt}1016\discretionary{/}{%
}{/}j\hspace{.1pt}\discretionary{.}{%
}{.}\hspace{.4pt}cag\hspace{.1pt}\discretionary{.}{%
}{.}\hspace{.4pt}2018\hspace{.1pt}\discretionary{.}{%
}{.}\hspace{.4pt}01\hspace{.1pt}\discretionary{.}{%
}{.}\hspace{.4pt}005}}}


\bibitem{rolff_interactive_2023}
\href{https://doi.org/10.1145/3607822.3618020}{T.~Rolff, K.~Li, J.~Hertel,
  S.~Schmidt, S.~Frintrop, and F.~Steinicke}.
\newblock \href{https://doi.org/10.1145/3607822.3618020}{Interactive
  {VRS}-{NeRF}: {Lightning} fast {Neural} {Radiance} {Field} {Rendering} for
  {Virtual} {Reality}}.
\newblock \href{https://doi.org/10.1145/3607822.3618020}{In {\em Proceedings of
  the 2023 {ACM} {Symposium} on {Spatial} {User} {Interaction}}},
  \href{https://doi.org/10.1145/3607822.3618020}{{SUI} '23},
  \href{https://doi.org/10.1145/3607822.3618020}{pp. 1--3}.
  \href{https://doi.org/10.1145/3607822.3618020}{Association for Computing
  Machinery}, \href{https://doi.org/10.1145/3607822.3618020}{New York, NY,
  USA}, \href{https://doi.org/10.1145/3607822.3618020}{Oct. 2023}.
  \href{https://doi.org/10.1145/3607822.3618020}
{doi: {{%
10\hspace{.1pt}\discretionary{.}{%
}{.}\hspace{.4pt}1145\discretionary{/}{%
}{/}3607822\hspace{.1pt}\discretionary{.}{%
}{.}\hspace{.4pt}3618020}}}


\bibitem{rowe_application_2019}
\href{https://doi.org/10.1007/s00261-019-02154-5}{S.~P. Rowe, L.~C. Chu, A.~R.
  Meyer, M.~A. Gorin, and E.~K. Fishman}.
\newblock \href{https://doi.org/10.1007/s00261-019-02154-5}{The application of
  cinematic rendering to {CT} evaluation of upper tract urothelial tumors:
  principles and practice}.
\newblock \href{https://doi.org/10.1007/s00261-019-02154-5}{{\em Abdominal
  Radiology}},
  \href{https://doi.org/10.1007/s00261-019-02154-5}{44(12):3886--3892},
  \href{https://doi.org/10.1007/s00261-019-02154-5}{Dec. 2019}.
  \href{https://doi.org/10.1007/s00261-019-02154-5}
{doi: {{%
10\hspace{.1pt}\discretionary{.}{%
}{.}\hspace{.4pt}1007\discretionary{/}{%
}{/}s00261\discretionary{%
}{-}{-}019\discretionary{%
}{-}{-}02154\discretionary{%
}{-}{-}5}}}


\bibitem{schonberger_structure--motion_2016}
\href{https://doi.org/10.1109/CVPR.2016.445}{J.~L. Schönberger and J.-M.
  Frahm}.
\newblock \href{https://doi.org/10.1109/CVPR.2016.445}{Structure-from-{Motion}
  {Revisited}}.
\newblock \href{https://doi.org/10.1109/CVPR.2016.445}{In {\em 2016 {IEEE}
  {Conference} on {Computer} {Vision} and {Pattern} {Recognition} ({CVPR})}},
  \href{https://doi.org/10.1109/CVPR.2016.445}{pp. 4104--4113},
  \href{https://doi.org/10.1109/CVPR.2016.445}{June 2016}.
  \href{https://doi.org/10.1109/CVPR.2016.445}
{doi: {{%
10\hspace{.1pt}\discretionary{.}{%
}{.}\hspace{.4pt}1109\discretionary{/}{%
}{/}CVPR\hspace{.1pt}\discretionary{.}{%
}{.}\hspace{.4pt}2016\hspace{.1pt}\discretionary{.}{%
}{.}\hspace{.4pt}445}}}


\bibitem{serra_overcoming_2018}
\href{https://doi.org/10.48550/arXiv.1801.01423}{J.~Serrà, D.~Surís,
  M.~Miron, and A.~Karatzoglou}.
\newblock \href{https://doi.org/10.48550/arXiv.1801.01423}{Overcoming
  catastrophic forgetting with hard attention to the task},
  \href{https://doi.org/10.48550/arXiv.1801.01423}{May 2018}.
  \href{https://doi.org/10.48550/arXiv.1801.01423}
{doi: {{%
10\hspace{.1pt}\discretionary{.}{%
}{.}\hspace{.4pt}48550\discretionary{/}{%
}{/}arXiv\hspace{.1pt}\discretionary{.}{%
}{.}\hspace{.4pt}1801\hspace{.1pt}\discretionary{.}{%
}{.}\hspace{.4pt}01423}}}


\bibitem{taibo_immersive_2024}
\href{https://doi.org/10.1109/VR58804.2024.00123}{J.~Taibo and J.~A.
  Iglesias-Guitian}.
\newblock \href{https://doi.org/10.1109/VR58804.2024.00123}{Immersive {3D}
  {Medical} {Visualization} in {Virtual} {Reality} using {Stereoscopic}
  {Volumetric} {Path} {Tracing}}.
\newblock \href{https://doi.org/10.1109/VR58804.2024.00123}{In {\em 2024 {IEEE}
  {Conference} {Virtual} {Reality} and {3D} {User} {Interfaces} ({VR})}},
  \href{https://doi.org/10.1109/VR58804.2024.00123}{pp. 1044--1053},
  \href{https://doi.org/10.1109/VR58804.2024.00123}{Mar. 2024}.
  \href{https://doi.org/10.1109/VR58804.2024.00123}
{doi: {{%
10\hspace{.1pt}\discretionary{.}{%
}{.}\hspace{.4pt}1109\discretionary{/}{%
}{/}VR58804\hspace{.1pt}\discretionary{.}{%
}{.}\hspace{.4pt}2024\hspace{.1pt}\discretionary{.}{%
}{.}\hspace{.4pt}00123}}}


\bibitem{vazquez_representative_2008}
\href{https://doi.org/10.1007/978-3-540-85412-8_10}{P.-P. Vázquez,
  E.~Monclús, and I.~Navazo}.
\newblock \href{https://doi.org/10.1007/978-3-540-85412-8_10}{Representative
  {Views} and {Paths} for {Volume} {Models}}.
\newblock \href{https://doi.org/10.1007/978-3-540-85412-8_10}{In A.~Butz,
  B.~Fisher, A.~Krüger, P.~Olivier, and M.~Christie, eds., {\em Smart
  {Graphics}}}, \href{https://doi.org/10.1007/978-3-540-85412-8_10}{pp.
  106--117}. \href{https://doi.org/10.1007/978-3-540-85412-8_10}{Springer},
  \href{https://doi.org/10.1007/978-3-540-85412-8_10}{Berlin, Heidelberg},
  \href{https://doi.org/10.1007/978-3-540-85412-8_10}{2008}.
  \href{https://doi.org/10.1007/978-3-540-85412-8_10}
{doi: {{%
10\hspace{.1pt}\discretionary{.}{%
}{.}\hspace{.4pt}1007\discretionary{/}{%
}{/}978\discretionary{%
}{-}{-}3\discretionary{%
}{-}{-}540\discretionary{%
}{-}{-}85412\discretionary{%
}{-}{-}8\_10}}}


\bibitem{walsh_imaging_2021}
\href{https://doi.org/10.1038/s41592-021-01317-x}{C.~L. Walsh, P.~Tafforeau,
  W.~L. Wagner, D.~J. Jafree, A.~Bellier, C.~Werlein, M.~P. Kühnel, E.~Boller,
  S.~Walker-Samuel, J.~L. Robertus, D.~A. Long, J.~Jacob, S.~Marussi, E.~Brown,
  N.~Holroyd, D.~D. Jonigk, M.~Ackermann, and P.~D. Lee}.
\newblock \href{https://doi.org/10.1038/s41592-021-01317-x}{Imaging intact
  human organs with local resolution of cellular structures using hierarchical
  phase-contrast tomography}.
\newblock \href{https://doi.org/10.1038/s41592-021-01317-x}{{\em Nature
  Methods}},
  \href{https://doi.org/10.1038/s41592-021-01317-x}{18(12):1532--1541},
  \href{https://doi.org/10.1038/s41592-021-01317-x}{Dec. 2021}.
  \href{https://doi.org/10.1038/s41592-021-01317-x}
{doi: {{%
10\hspace{.1pt}\discretionary{.}{%
}{.}\hspace{.4pt}1038\discretionary{/}{%
}{/}s41592\discretionary{%
}{-}{-}021\discretionary{%
}{-}{-}01317\discretionary{%
}{-}{-}x}}}


\bibitem{wang_image_2004}
\href{https://doi.org/10.1109/TIP.2003.819861}{Z.~Wang, A.~Bovik, H.~Sheikh,
  and E.~Simoncelli}.
\newblock \href{https://doi.org/10.1109/TIP.2003.819861}{Image quality
  assessment: from error visibility to structural similarity}.
\newblock \href{https://doi.org/10.1109/TIP.2003.819861}{{\em IEEE Transactions
  on Image Processing}},
  \href{https://doi.org/10.1109/TIP.2003.819861}{13(4):600--612},
  \href{https://doi.org/10.1109/TIP.2003.819861}{Apr. 2004}.
  \href{https://doi.org/10.1109/TIP.2003.819861}
{doi: {{%
10\hspace{.1pt}\discretionary{.}{%
}{.}\hspace{.4pt}1109\discretionary{/}{%
}{/}TIP\hspace{.1pt}\discretionary{.}{%
}{.}\hspace{.4pt}2003\hspace{.1pt}\discretionary{.}{%
}{.}\hspace{.4pt}819861}}}


\bibitem{wasserthal_totalsegmentator_2023}
\href{https://doi.org/10.1148/ryai.230024}{J.~Wasserthal, H.-C. Breit, M.~T.
  Meyer, M.~Pradella, D.~Hinck, A.~W. Sauter, T.~Heye, D.~T. Boll, J.~Cyriac,
  S.~Yang, M.~Bach, and M.~Segeroth}.
\newblock \href{https://doi.org/10.1148/ryai.230024}{{TotalSegmentator}:
  {Robust} {Segmentation} of 104 {Anatomic} {Structures} in {CT} {Images}}.
\newblock \href{https://doi.org/10.1148/ryai.230024}{{\em Radiology: Artificial
  Intelligence}}, \href{https://doi.org/10.1148/ryai.230024}{5(5):e230024},
  \href{https://doi.org/10.1148/ryai.230024}{Sept. 2023}.
  \href{https://doi.org/10.1148/ryai.230024}
{doi: {{%
10\hspace{.1pt}\discretionary{.}{%
}{.}\hspace{.4pt}1148\discretionary{/}{%
}{/}ryai\hspace{.1pt}\discretionary{.}{%
}{.}\hspace{.4pt}230024}}}


\bibitem{wheeler_virtual_2018}
\href{https://doi.org/10.1049/htl.2018.5064}{G.~Wheeler, S.~Deng, N.~Toussaint,
  K.~Pushparajah, J.~A. Schnabel, J.~M. Simpson, and A.~Gomez}.
\newblock \href{https://doi.org/10.1049/htl.2018.5064}{Virtual interaction and
  visualisation of {3D} medical imaging data with {VTK} and {Unity}}.
\newblock \href{https://doi.org/10.1049/htl.2018.5064}{{\em Healthcare
  Technology Letters}},
  \href{https://doi.org/10.1049/htl.2018.5064}{5(5):148--153},
  \href{https://doi.org/10.1049/htl.2018.5064}{Oct. 2018}.
  \href{https://doi.org/10.1049/htl.2018.5064}
{doi: {{%
10\hspace{.1pt}\discretionary{.}{%
}{.}\hspace{.4pt}1049\discretionary{/}{%
}{/}htl\hspace{.1pt}\discretionary{.}{%
}{.}\hspace{.4pt}2018\hspace{.1pt}\discretionary{.}{%
}{.}\hspace{.4pt}5064}}}


\bibitem{xie2024surgicalGaussiandeformable3dGaussians}
\href{https://arxiv.org/abs/2407.05023}{W.~Xie, J.~Yao, X.~Cao, Q.~Lin,
  Z.~Tang, X.~Dong, and X.~Guo}.
\newblock \href{https://arxiv.org/abs/2407.05023}{Surgicalgaussian: Deformable
  3d gaussians for high-fidelity surgical scene reconstruction},
  \href{https://arxiv.org/abs/2407.05023}{2024}.

\bibitem{yang_depth_2024}
\href{https://doi.org/10.1109/CVPR52733.2024.00987}{L.~Yang, B.~Kang, Z.~Huang,
  X.~Xu, J.~Feng, and H.~Zhao}.
\newblock \href{https://doi.org/10.1109/CVPR52733.2024.00987}{Depth {Anything}:
  {Unleashing} the {Power} of {Large}-{Scale} {Unlabeled} {Data}}.
\newblock \href{https://doi.org/10.1109/CVPR52733.2024.00987}{In {\em 2024
  {IEEE}/{CVF} {Conference} on {Computer} {Vision} and {Pattern} {Recognition}
  ({CVPR})}}, \href{https://doi.org/10.1109/CVPR52733.2024.00987}{pp.
  10371--10381}, \href{https://doi.org/10.1109/CVPR52733.2024.00987}{June
  2024}.
\newblock \href{https://doi.org/10.1109/CVPR52733.2024.00987}{ISSN: 2575-7075}.
  \href{https://doi.org/10.1109/CVPR52733.2024.00987}
{doi: {{%
10\hspace{.1pt}\discretionary{.}{%
}{.}\hspace{.4pt}1109\discretionary{/}{%
}{/}CVPR52733\hspace{.1pt}\discretionary{.}{%
}{.}\hspace{.4pt}2024\hspace{.1pt}\discretionary{.}{%
}{.}\hspace{.4pt}00987}}}


\bibitem{yu_mip-splatting_2023}
\href{https://doi.org/10.48550/arXiv.2311.16493}{Z.~Yu, A.~Chen, B.~Huang,
  T.~Sattler, and A.~Geiger}.
\newblock \href{https://doi.org/10.48550/arXiv.2311.16493}{Mip-{Splatting}:
  {Alias}-free {3D} {Gaussian} {Splatting}},
  \href{https://doi.org/10.48550/arXiv.2311.16493}{Nov. 2023}.
  \href{https://doi.org/10.48550/arXiv.2311.16493}
{doi: {{%
10\hspace{.1pt}\discretionary{.}{%
}{.}\hspace{.4pt}48550\discretionary{/}{%
}{/}arXiv\hspace{.1pt}\discretionary{.}{%
}{.}\hspace{.4pt}2311\hspace{.1pt}\discretionary{.}{%
}{.}\hspace{.4pt}16493}}}


\bibitem{zhang_volume_2011}
\href{https://doi.org/10.1007/s10278-010-9321-6}{Q.~Zhang, R.~Eagleson, and
  T.~M. Peters}.
\newblock \href{https://doi.org/10.1007/s10278-010-9321-6}{Volume
  {Visualization}: {A} {Technical} {Overview} with a {Focus} on {Medical}
  {Applications}}.
\newblock \href{https://doi.org/10.1007/s10278-010-9321-6}{{\em Journal of
  Digital Imaging}},
  \href{https://doi.org/10.1007/s10278-010-9321-6}{24(4):640--664},
  \href{https://doi.org/10.1007/s10278-010-9321-6}{Aug. 2011}.
  \href{https://doi.org/10.1007/s10278-010-9321-6}
{doi: {{%
10\hspace{.1pt}\discretionary{.}{%
}{.}\hspace{.4pt}1007\discretionary{/}{%
}{/}s10278\discretionary{%
}{-}{-}010\discretionary{%
}{-}{-}9321\discretionary{%
}{-}{-}6}}}


\bibitem{zhang_unreasonable_2018}
\href{https://doi.org/10.1109/CVPR.2018.00068}{R.~Zhang, P.~Isola, A.~A. Efros,
  E.~Shechtman, and O.~Wang}.
\newblock \href{https://doi.org/10.1109/CVPR.2018.00068}{The {Unreasonable}
  {Effectiveness} of {Deep} {Features} as a {Perceptual} {Metric}}.
\newblock \href{https://doi.org/10.1109/CVPR.2018.00068}{In {\em 2018
  {IEEE}/{CVF} {Conference} on {Computer} {Vision} and {Pattern}
  {Recognition}}}, \href{https://doi.org/10.1109/CVPR.2018.00068}{pp.
  586--595}, \href{https://doi.org/10.1109/CVPR.2018.00068}{June 2018}.
\newblock \href{https://doi.org/10.1109/CVPR.2018.00068}{ISSN: 2575-7075}.
  \href{https://doi.org/10.1109/CVPR.2018.00068}
{doi: {{%
10\hspace{.1pt}\discretionary{.}{%
}{.}\hspace{.4pt}1109\discretionary{/}{%
}{/}CVPR\hspace{.1pt}\discretionary{.}{%
}{.}\hspace{.4pt}2018\hspace{.1pt}\discretionary{.}{%
}{.}\hspace{.4pt}00068}}}


\bibitem{zhao2024hfgs4dGaussiansplatting}
\href{https://arxiv.org/abs/2405.17872}{H.~Zhao, X.~Zhao, L.~Zhu, W.~Zheng, and
  Y.~Xu}.
\newblock \href{https://arxiv.org/abs/2405.17872}{Hfgs: 4d gaussian splatting
  with emphasis on spatial and temporal high-frequency components for
  endoscopic scene reconstruction},
  \href{https://arxiv.org/abs/2405.17872}{2024}.

\bibitem{zhou_review_2022}
\href{https://doi.org/10.34133/2022/9840519}{L.~Zhou, M.~Fan, C.~Hansen, C.~R.
  Johnson, and D.~Weiskopf}.
\newblock \href{https://doi.org/10.34133/2022/9840519}{A {Review} of
  {Three}-{Dimensional} {Medical} {Image} {Visualization}}.
\newblock \href{https://doi.org/10.34133/2022/9840519}{{\em Health Data
  Science}}, \href{https://doi.org/10.34133/2022/9840519}{2022:9840519},
  \href{https://doi.org/10.34133/2022/9840519}{Apr. 2022}.
\newblock \href{https://doi.org/10.34133/2022/9840519}{Publisher: American
  Association for the Advancement of Science}.
  \href{https://doi.org/10.34133/2022/9840519}
{doi: {{%
10\hspace{.1pt}\discretionary{.}{%
}{.}\hspace{.4pt}34133\discretionary{/}{%
}{/}2022\discretionary{/}{%
}{/}9840519}}}


\bibitem{tcga-hnsc}
\href{https://doi.org/10.7937/K9/TCIA.2016.LXKQ47MS}{M.~L. Zuley, R.~Jarosz,
  S.~Kirk, Y.~Lee, R.~Colen, K.~Garcia, D.~Delbeke, M.~Pham, P.~Nagy,
  G.~Sevinc, M.~Goldsmith, S.~Khan, J.~M. Net, F.~R. Lucchesi, and N.~D.
  Aredes}.
\newblock \href{https://doi.org/10.7937/K9/TCIA.2016.LXKQ47MS}{The cancer
  genome atlas head-neck squamous cell carcinoma collection (tcga-hnsc)},
  \href{https://doi.org/10.7937/K9/TCIA.2016.LXKQ47MS}{2016}.
  \href{https://doi.org/10.7937/K9/TCIA.2016.LXKQ47MS}
{doi: {{%
10\hspace{.1pt}\discretionary{.}{%
}{.}\hspace{.4pt}7937\discretionary{/}{%
}{/}K9\discretionary{/}{%
}{/}TCIA\hspace{.1pt}\discretionary{.}{%
}{.}\hspace{.4pt}2016\hspace{.1pt}\discretionary{.}{%
}{.}\hspace{.4pt}LXKQ47MS}}}


\bibitem{zwicker_ewa_2001}
\href{https://doi.org/10.1109/VISUAL.2001.964490}{M.~Zwicker, H.~Pfister,
  J.~van Baar, and M.~Gross}.
\newblock \href{https://doi.org/10.1109/VISUAL.2001.964490}{{EWA} volume
  splatting}.
\newblock \href{https://doi.org/10.1109/VISUAL.2001.964490}{In {\em Proceedings
  {Visualization}, 2001. {VIS} '01.}},
  \href{https://doi.org/10.1109/VISUAL.2001.964490}{pp. 29--538},
  \href{https://doi.org/10.1109/VISUAL.2001.964490}{Oct. 2001}.
  \href{https://doi.org/10.1109/VISUAL.2001.964490}
{doi: {{%
10\hspace{.1pt}\discretionary{.}{%
}{.}\hspace{.4pt}1109\discretionary{/}{%
}{/}VISUAL\hspace{.1pt}\discretionary{.}{%
}{.}\hspace{.4pt}2001\hspace{.1pt}\discretionary{.}{%
}{.}\hspace{.4pt}964490}}}


\end{thebibliography}


\makeatletter
\renewcommand \thesection{S\@arabic\c@section}
\renewcommand\thetable{S\@arabic\c@table}
\renewcommand \thefigure{S\@arabic\c@figure}
\makeatother

\newpage

\begin{figure*}[ht!]
    \centering
    \includegraphics[width=0.9\linewidth]{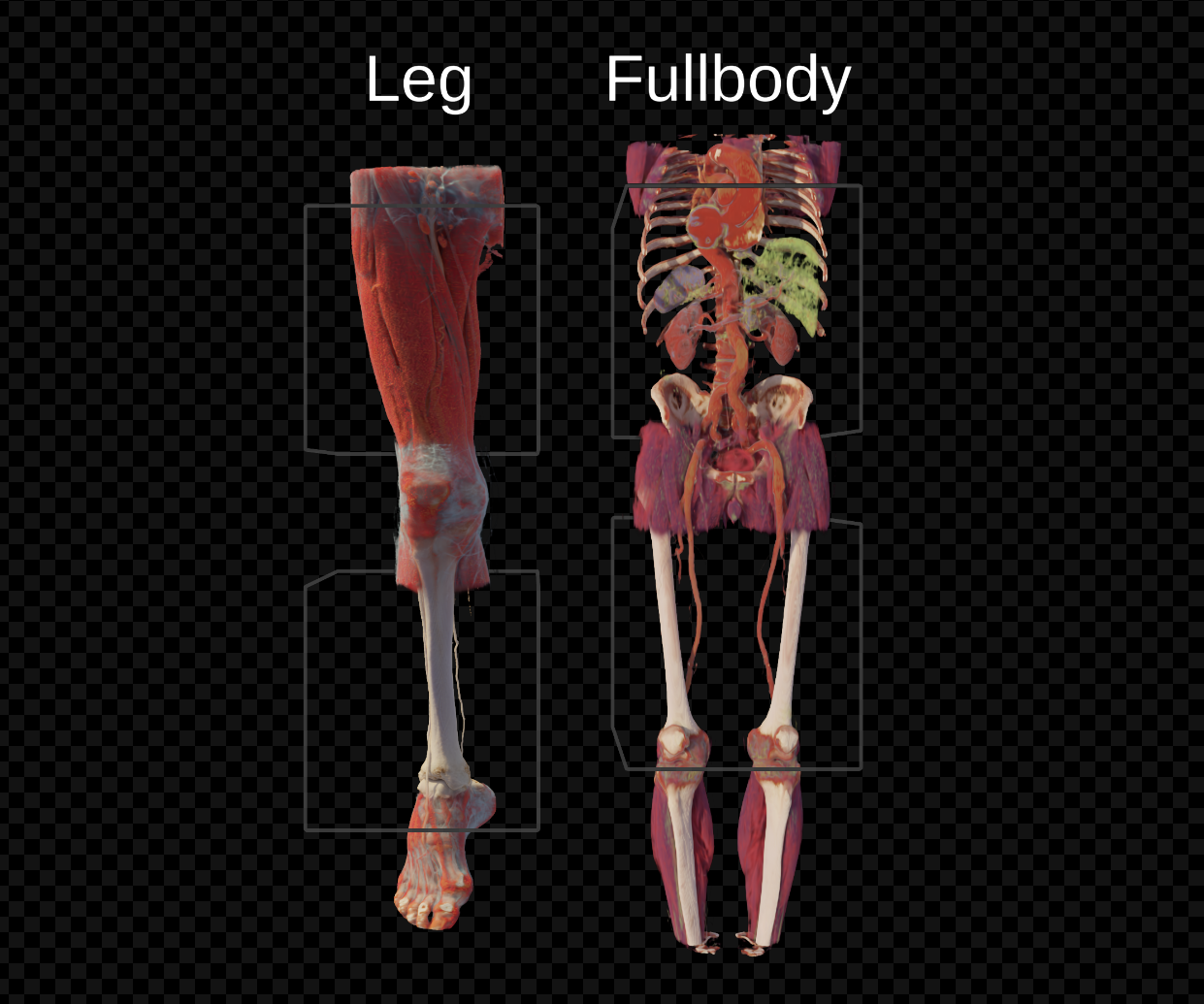}
    \caption{Exemplary scene with layers cut in leg and fullbody scenes, as seen from within the game engine.}
    \label{fig:cutting}
\end{figure*}

\section{Supplementary}
\label{sec:supplementary_materials}

Supplemental materials, generated training data, and trained models are available at \url{https://osf.io/tuwh5/}. Code is available at \url{https://github.com/roth-hex-lab/Multi-Layer-Anatomy-GS-Unity-Rendering}.
The original CT volumes used in this work are available from their respective authors.

\paragraph{Data generation}
To ensure noise-free images in our ground truth data, the path tracer uses 4096 samples per pixel and allows up to 100 light bounces. Rendering begins with full-object views and progresses to close-up shots of different anatomical parts from various angles, ensuring a comprehensive representation of the anatomical structures. All images, both anatomical and synthetic, include transparent backgrounds. Note that in the bone scenes, some blood vessels are visible due to their density being similar to that of bony soft tissue.

\paragraph{Layering}
Layered reconstruction generally works well, even at default GS settings; see metrics in \autoref{tab:qualitycomparison2}. Lower layers often reconstruct slightly worse than upper layers. This is possibly due to general increases in visual complexity in upper layers, as well as fixed Gaussians from lower layers that would not usually be in upper layers. This is best avoided in input data generation but can sometimes be cumbersome. 

\paragraph{Compression}
We evaluate the compression performance of our approach against the commonly used DICOM and compressed Nifti file formats, see \autoref{tab:filesizecomparison2}. Note that while the internal data structure used by the path tracer is more compact than the on-file representation, it is generally less portable and not typically stored.

\begin{figure*}[t!]
    \centering
    \includegraphics[width=\linewidth]{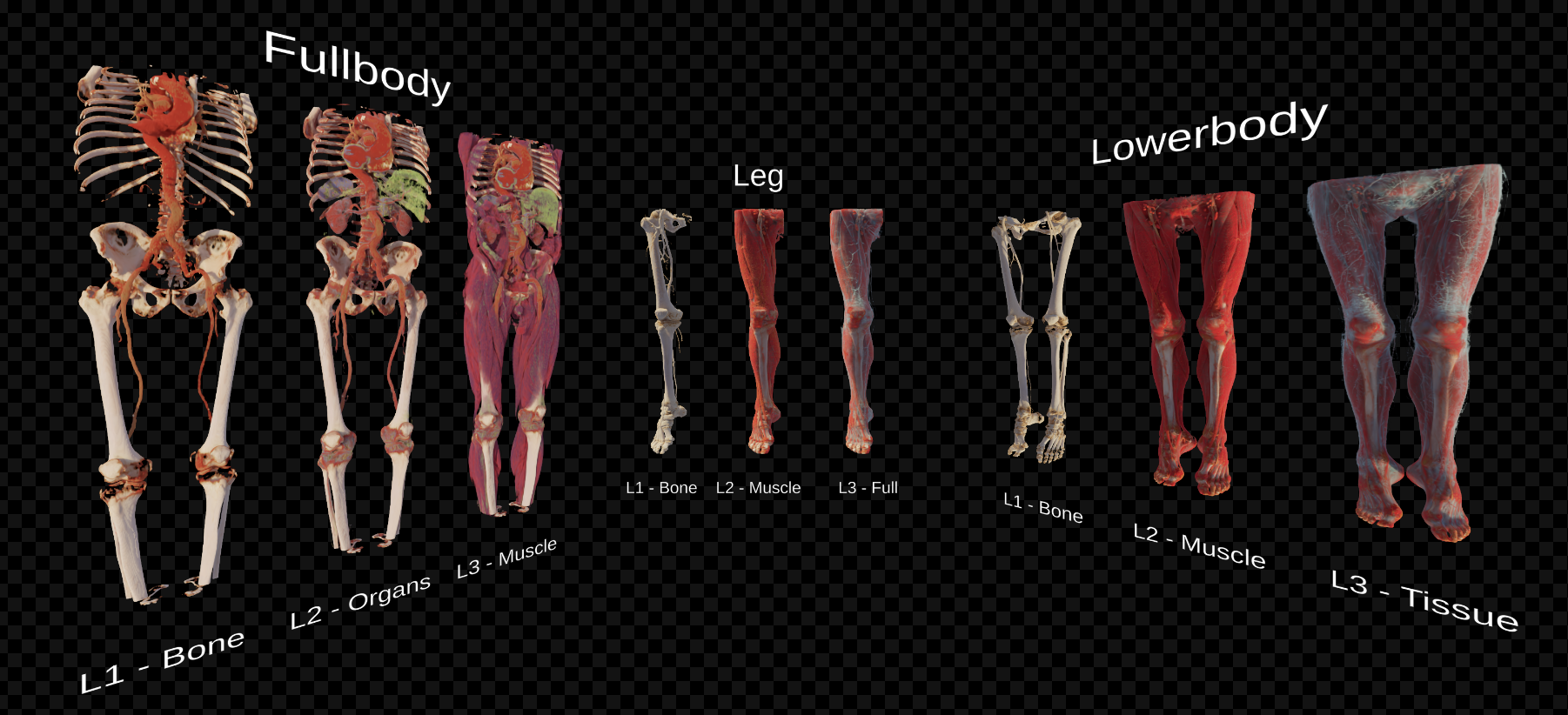}
    \caption{Overview of all CT data scenes used in our anatomy dataset, as seen from within the game engine.}
    \label{fig:anatomy_overview}
\end{figure*}

\begin{figure*}[t!]
    \centering
    \includegraphics[width=\linewidth]{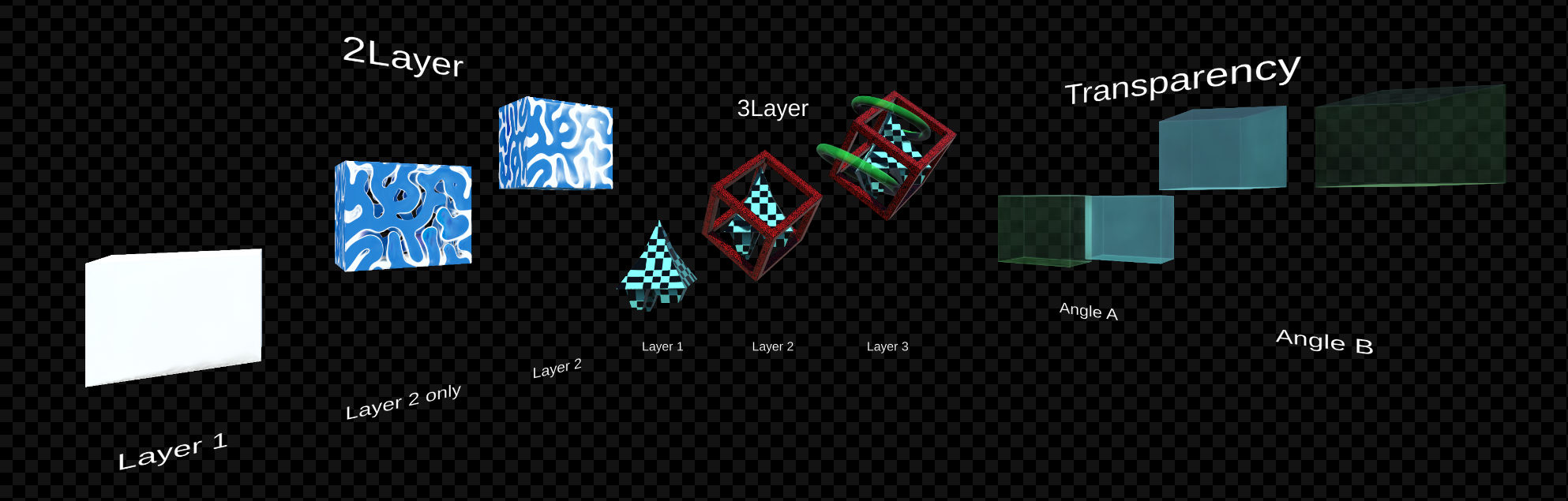}
    \caption{Overview of all scenes used in our synthetic dataset, as seen from within the game engine. Some objects can show visual artifacts from previously unseen viewing angles due to the limited variety in camera orientations towards the object in our synthetic datasets.}
    \label{fig:synth_overview}
\end{figure*}

\paragraph{Renderings}
We include additional overview renderings as they would be seen from within the game engine for all anatomy scenes~(\autoref{fig:anatomy_overview}), synthetic scenes~(\autoref{fig:synth_overview}), and selective layer cutting~(\autoref{fig:cutting}).

\begin{table*}[t!]
\centering
\caption{Comparison of on-disk sizes between different medical volumes before and after training and compression.}
\begin{tabularx}{\linewidth}{@{}lXXXXXX@{}}
\toprule
& \multicolumn{2}{c}{Leg} & \multicolumn{2}{c}{Lower Body} & \multicolumn{2}{c}{Fullbody} \\ \cmidrule(l){2-3} \cmidrule(l){4-5} \cmidrule(l){6-7}
& Normal & HQ Recon. & Normal & HQ Recon. & Normal & HQ Recon. \\ \midrule
Original Volume (DICOM)                    & 304.7   & 304.7      & 1585.1     & 1585.1    & 87.4    & 87.4 \\
Compressed (.nii.gz)                       & 73.5    & 73.5       & 730.8      & 730.8     & 75.5    & 75.5 \\
Reconstructed PLY file                     & 32.4    & 62.2       & 22.2       & 45.3      & 47.6    & 92.2 \\
Unity Comp. (High Quality)                 & 12.0    & 22.9       & 8.3        & 16.7      & 17.6    & 33.9 \\
Unity Comp. (Low Filesize)                 & 5.6     & 9.2        & 4.4        & 7.1       & 7.4     & 12.8 \\ \midrule
Original compressed up to                 & 98.16\% &  96.98\%   & 99.72\%    & 99.55\%   & 91.53\% & 85.35\% \\ \bottomrule
\end{tabularx}
\label{tab:filesizecomparison2}
\end{table*}

\begin{table*}[t!]
\centering
\caption{Comparison of reconstruction quality at default GS settings before import to Unity and compression.}
\begin{tabularx}{\linewidth}{@{}lXXXXXXXXX@{}}
\toprule
& \multicolumn{3}{c}{Leg} & \multicolumn{3}{c}{Lower Body} & \multicolumn{3}{c}{Fullbody} \\ \cmidrule(l){2-4} \cmidrule(l){5-7} \cmidrule(l){8-10}
    Layer        & 1      & 2      & 3      & 1      & 2      & 3      & 1      &  2     & 3      \\ \midrule
PSNR ↑           & 38.575 & 35.851 & 30.367 & 38.165 & 37.267 & 31.028 & 36.187 & 37.370 & 33.559 \\
SSIM ↑           & 0.990  & 0.971  & 0.957  & 0.991  & 0.968  & 0.940  & 0.986  & 0.972  & 0.970  \\
LPIPS ↓          & 0.023  & 0.042  & 0.064  & 0.022  & 0.051  & 0.082  & 0.024  & 0.056  & 0.045  \\ \bottomrule
\end{tabularx}
\label{tab:qualitycomparison2}
\end{table*}

\end{document}